\documentclass[11pt,eqsecnum]{article}
\usepackage{amsmath,amsfonts,amssymb,latexsym,theorem,mathrsfs,latexsym,cite,setspace,
comment,color,multirow,graphicx}
\usepackage{hyperref}
\pdfoutput=1

\newcommand{\ma}[1]{\mbox{$\mathcal{#1}$}}
\newcommand{\mas}[1]{\mbox{$\mathscr{#1}$}}

\newcommand{\D}{{\rm d}}

\begin{document}

\begin{titlepage}

\begin{flushright}
{
\small 
YITP-20-134\\
\today
}
\end{flushright}
\vspace{1cm}

\begin{center}
%---------- title ----------%
{\LARGE \bf
\begin{spacing}{1}
Static spacetimes haunted by a phantom scalar field III: 
\\ \vspace{0.2cm}
asymptotically (A)dS solutions
\end{spacing}
}
\end{center}
\vspace{.5cm}

\begin{center}
%---------- author ----------%
{\large \bf
Masato Nozawa
} \\

\vskip 1cm
{\it
Center for Gravitational Physics, Yukawa Institute for Theoretical Physics, \\
 Kyoto University, Kyoto 606-8502, Japan.}\\
\texttt{masato.nozawa@yukawa.kyoto-u.ac.jp}

\end{center}

\vspace{.5cm}

%---------- abstract ----------%
\begin{abstract}
The static and spherically symmetric solutions in $n(\ge 4)$-dimensional Einstein-phantom-scalar system fall into three family: (i) the Fisher solution, (ii) the Ellis-Gibbons solution, and (iii) the Ellis-Bronnikov solution. We exploit these solutions as seed to generate a bunch of corresponding asymptotically (A)dS spacetimes, at the price of introducing the potential of the scalar field. Despite that the potentials are different for each solution, each potential is expressed in terms of the superpotential as in supergravity. 
We discuss the global structure of these solutions in detail and spell out the domain of parameters under which each solution represents a black hole/wormhole. The Ellis-Bronnikov class of solutions presents novel examples of spherical traversable wormholes that interpolate two different (A)dS critical points of the (super)potential.
\end{abstract}

\vspace{.5cm}

\setcounter{footnote}{0}

\end{titlepage}

\setcounter{tocdepth}{2}
\tableofcontents

\newpage

%==============================================%
%<<<<<<<<<<<< SECTION I Introduction >>>>>>>>>>>>>>%
%==============================================%
\section{Introduction}

The advent of the AdS/CFT correspondence and gauge/gravity duality \cite{Maldacena:1997re} has sparkled a considerable interest in the anti-de Sitter (AdS) spacetimes. 
Asymptotically AdS black holes are expected to describe the dual thermal states of a boundary conformal field theory.  Despite the increasing importance of asymptotically AdS solutions, 
the exact solutions are fairly elusive, contrary to the asymptotically flat case. 
In the latter case, we have a powerful mechanism that allows us to generate new solutions 
based on the symmetry for the nonlinear sigma model of the scalar fields. 
In the context of supergravity, AdS vacua appear as the critical point of the scalar field potential.
Any studies of seeking useful algorithms for constructing new solutions in AdS spaces 
are hampered by the fact that the symmetry of the sigma model is partly or completely broken
by the scalar potential, or more simply by the cosmological constant \cite{Klemm:2015uba}. This circumstance makes it exceedingly difficult to systematically construct gravitational solutions of physical interest in AdS. 

In the spirit of holographic entanglement and quantum teleportation, the past decade has witnessed a resurging interest for the wormholes in AdS \cite{Maldacena:2001kr,Maldacena:2004rf,Gao:2016bin,Maldacena:2017axo,vanBreukelen:2017dul,Maldacena:2018lmt}. Wormholes describe the neck structure of spacetimes bridging  two separate universes \cite{visser,Lobo:2007zb}. 
Much effort has been devoted to the global structure \cite{Morris:1988cz,Hochberg:1998ha,Kim:2013tsa}, stability \cite{Novikov:2009vn,Nandi:2016ccg,Cremona:2018wkj},  
astrophysical signature \cite{Kardashev:2006nj}, and possible time travels \cite{Morris:1988tu} in wormhole spacetimes. 
The most prominent property of the wormholes is that they disobey the standard energy conditions \cite{Morris:1988tu,Friedman:1993ty,Galloway}. In the classical perspective, the requirement of suitable energy conditions appears to be physically sensible~\cite{Hawking:1973uf,Maeda:2018hqu}. For instance, the dominant energy condition is called for in the proof of the positive energy theorem~\cite{Schon:1979rg,SchonYau,Witten:1981mf}, which guarantees the stability of the ground state. 
In the aforementioned holographic context, wormholes are sustained instead by quantum fields. Recently, the new nonlocal concept of the quantum null energy condition has been proposed 
\cite{Bousso:2015mna,Bousso:2015wca}, and its possible violation has been discussed extensively \cite{Ishibashi:2018ern}. Putting this mainstream aside, 
it is fully encouraging to clarify the properties of wormhole spacetimes in AdS within the classical regime. 
So far, not so many fully-fledged examples of AdS wormhole solutions  are available in the literature. 
Some nontrivial examples are the solution obtained by the cut-and-paste technique \cite{Lemos:2003jb,Lemos:2004vs}, the higher curvature solution \cite{Maeda:2008nz}, the dynamical solution \cite{Maeda:2012fr} and the asymptotically locally AdS solution \cite{Anabalon:2018rzq}.

Combining the above two motivations together, we try in this paper to find a prescription for constructing exact asymptotically (A)dS solutions with a phantom field. 
We consider the static and spherically symmetric solutions to the Einstein-phantom scalar system with a potential.  We contrive a simple ansatz that gives rise to the asymptotically (A)dS spacetimes 
from the asymptotically flat seed solutions in the Einstein-phantom scalar system without a potential. 
In our previous paper \cite{paperI}, we generalized the illustrious work by Ellis  \cite{Ellis1973} 
and demonstrated that the static and spherically symmetric solutions to the $n$-dimensional Einstein-phantom scalar system are systematically classified into three different family of solutions (see \cite{paperII} for the dilatonic charged solutions): 
the Fisher solution \cite{Fisher:1948yn}, the Ellis-Gibbons solution \cite{Ellis1973,Gibbons:2003yj,Gibbons:2017jzk}, and the Ellis-Bronnikov solution  \cite{Ellis1973,Bronnikov1973}. 
Among these, only the Fisher class of solutions is allowed to exist in a system with a conventional scalar field, and other two solutions are intrinsic to the phantom case. The Fisher solution has been rediscovered many times by  Bergmann and Leipnik \cite{Bergmann:1957zza}, Buchdahl \cite{Buchdahl:1959nk},  Janis, Newman and Winicour \cite{jnw1968}, Ellis \cite{Ellis1973}, Bronnikov \cite{Bronnikov1973} and Wyman \cite{Wyman:1981bd}.  
The Ellis-Gibbons class  is sometimes called the ``exponential metric''  and can be generalized to be a multi-center solution \cite{Gibbons:2003yj,Gibbons:2017jzk}. As established in \cite{paperI}, the Fisher and the Ellis-Gibbons solutions always suffer from the naked scalar/parallelly propagated (p.p) curvature singularity in the entire parameter region (excluding the Schwarzschild case). In contrast, the Ellis-Bronnikov solution is regular and describes the famous wormhole geometry connecting two asymptotically flat regions. It is then natural to ask if these solutions can be embedded into (A)dS spaces.

This paper devises a simple recipe to generate solutions in (A)dS space from the asymptotically flat solutions. Our algorithm to generate new solutions is explained as follows. 
First we write the corresponding spherical solutions in the asymptotically flat spacetimes in terms of the isotropic coordinate. Then, we insert the time-dependence for the spacetime in such a way that the solution asymptotically approaches to the de Sitter (dS) universe, and that the solution is invariant under an additional scaling symmetry. The obtained metric does not solve the field equations in the original Einstein-phantom scalar system. Nevertheless, the time-dependence can be precisely offset by the introduction of the appropriate scalar potential. Lastly, we can cast the metric into the static patch of the dS universe by suitable coordinate transformations. The Hubble parameter of the solution is then Wick-rotated to obtain the asymptotically AdS spacetimes. Furthermore, the final expression of the metric can be used to make yet another ansatz to yield new asymptotically (A)dS solutions.  This simple-minded scheme turns out to be of extreme help to economically access the novel exact solutions.

In the foregoing procedure, the potential of the phantom scalar field is derived, rather than given beforehand. 
For each family of the solutions, the derived potentials are considerably different in appearance. 
Although these scalar  potentials are determined a posteriori, it deserves to stress that each potential is given in terms of the ``superpotential.'' The existence of the superpotential is quite nontrivial and indicative of the profound supergravity origin. 

Given the exact solutions at hand, we will next elaborate on the global structure of these solutions in detail. We show that the Fisher and the Ellis-Gibbons solutions in (A)dS may admit a black hole horizon which covers the central scalar/p.p curvature singularity. This property is not shared by the asymptotically flat counterparts. We further reveal that the (A)dS generalizations of the Ellis-Bronnikov solutions would describe the regular traversable wormhole spacetimes. These wormhole solutions differ in several respects from those obtained in the literature \cite{Lemos:2003jb,Lemos:2004vs,Maeda:2008nz,Maeda:2012fr,Anabalon:2018rzq}. 
Our exact, static and spherically symmetric wormhole solution is smooth everywhere (i.e., 
free of any distributional shells) and admits geodesic completeness. Further, our solution asymptotically approaches to the global, rather than local, AdS spacetime with $\mas I\simeq \mathbb R\times S^{n-2}$, corresponding to the critical point of the potential origin. After the maximal extension, the other side of AdS vacuum corresponds to the different critical point of the scalar potential. Namely, our solution interpolates two different vacua, which is reminiscent of the fundamental property of the soliton. 
Our new wormhole solutions would therefore be a valuable cornerstone in the context of holography.

The construction of the present paper is as follows. 
In the next section, we encapsulate the static and spherically symmetric solutions to 
the Einstein-phantom-scalar system, and illustrate the construction procedure for (A)dS solutions. In the following three sections \ref{sec:Fisher}, \ref{sec:Gibbons}, \ref{sec:EG}, we derive a bunch of new asymptotically (A)dS solutions corresponding to the generalization of the Fisher, Gibbons and Ellis-Bronnikov class respectively, and explore their causal structures. 
We summarize the conclusions of the present paper in section \ref{sec:conclusion}. 
Appendix \ref{app:curv} summarizes curvature formula and 
appendix \ref{Gibbons:FLRW} presents the extension of the Gibbons solution into the Friedman-Lema\^itre-Robertson-Walker (FLRW) universe.

Our basic notations follow~\cite{wald}.
The conventions of curvature tensors are 
$[\nabla _\rho ,\nabla_\sigma]V^\mu ={R^\mu }_{\nu\rho\sigma}V^\nu$ 
and ${R}_{\mu \nu }={R^\rho }_{\mu \rho \nu }$.
The Lorentzian metric is taken to be the mostly plus sign, and 
Greek indices run over all spacetime indices. 
We denote the $n$-dimensional gravitational
constant to be $\kappa_n=8\pi G_n$ for brevity.

\bigskip\noindent{\bf Note added.} 
After the submission of the paper to arXiv, we were informed by 
Jinbo Yang that they recently derived the (A)dS Ellis-Bronnikov wormhole 
in four dimensions \cite{Huang:2020qmn}.  Their discussion has some overlaps in 
section \ref{sec:EG}.

\section{Einstein's gravity with a phantom scalar}
\label{sec:pre}

We begin this paper by a brief review of static and spherically symmetric solutions to 
the $n$-dimensional Einstein-phantom-scalar system. A comprehensive analysis can be found in our previous paper \cite{paperI}. Next, we explain the procedure to obtain the asymptotically (A)dS solutions.

\subsection{Massless scalar}

The Einstein-(phantom-)scalar field system is described by the action
\begin{align}
\label{}
S=\int \D ^n x\sqrt{-g} \left( \frac 1{2\kappa_n } R-\frac 12 \epsilon (\nabla \phi)^2 \right)\,, 
\end{align}
where $\epsilon=+1$ corresponds to the ordinary scalar field, whereas 
the $\epsilon=-1$ case represents the phantom scalar field meditating the repulsive force. 
We are primarily interested in the static metric with the spherical symmetry
\begin{align}
\label{metric}
\D s^2=-f_1(r)\D t^2+f_2(r) \D r^2+S^2(r) \D \Omega_{ n-2}^2 \,, \qquad 
\phi=\phi(r) \,,
\end{align}
where $  \D \Omega_{ n-2}^2=\gamma_{ij}(z)\D z^i \D z^j$ is the line element of the unit sphere in $(n-2)$-dimensions, and $S=S(r)$ is the areal radius. 
As demonstrated in \cite{paperI}, there appear three classes of solutions:  the Fisher solution
\cite{Fisher:1948yn}, the Ellis-Gibbons solution \cite{Ellis1973,Gibbons:2003yj,Gibbons:2017jzk}, and the Ellis-Bronnikov solution  \cite{Ellis1973,Bronnikov1973}. 
The classification scheme and the global structures of these solutions have been 
discussed in detail in our previous paper \cite{paperI}. 
 
In order to make the present paper to be self-contained, let us 
summarize the necessary ingredients to inquire the global structure for the metric of the form (\ref{metric}). 
To this aim, a crucial role is played by the 
affine-parameterized radial null geodesics for the metric (\ref{metric}), whose tangent vector is given by 
\begin{align}
\label{}
k^\mu =\frac{1}{f_1(r)}\left(\frac{\partial}{\partial t}\right)^\mu \pm \frac 1{\sqrt{f_1(r)f_2(r)}}
\left(\frac{\partial}{\partial r}\right)^\mu\,, \qquad k^\nu \nabla_\nu k^\mu=0 \,. 
\end{align}
The affine parameter $\lambda $ of the geodesics can then be extracted (modulo affine transformation) as 
\begin{align}
\label{affine}
\lambda = \pm \int ^r  {\sqrt{f_1(r)f_2(r)}}\D r \,. 
\end{align}

If the scalar polynomials of the curvature tensors become unboundedly large, 
one concludes the existence of a spacetime curvature singularity. Instead, the p.p curvature singularity is characterized by the divergence of the Riemann tensor components in a frame parallelly propagated along some curve~\cite{Hawking:1973uf}. The existence of the p.p curvature singularities cannot be captured solely by the curvature polynomials. 
For our present purpose, the double projection of the radial null geodesic tangent vector $k^\mu$ onto the Ricci tensor is of importance, giving
\begin{align}
\label{}
R_{\mu\nu}k^\mu k^\nu =(n-2) \frac{f_2f_1'S'+f_1 f_2'S'-2 f_1 f_2 S''}{2f_1^2 f_2 ^2 S} \,,
\end{align}
where the prime denotes the differentiation with respect to $r$. 
Taking the orthonormal frame $e^{\hat i}{}_i $ by 
$\gamma_{ij}(z)=\delta_{\hat i\hat j}e^{\hat i}{}_i e^{\hat j}{}_j$ and defining
$E^{\hat i}{}_\mu =S(r) e^{\hat i}{}_i (\D z^i)_\mu$, one finds  
$k^\mu \nabla_\mu E^{\hat i}{}_\nu=0$, i.e., this frame is  parallelly propagated
along the radial null geodesics with the tangent vector $k^\mu$. 
Then one has 
$R_{\mu\nu\rho\sigma}k^\mu E_{\hat i}{}^\nu k^\rho E_{\hat j}{}^\sigma=(n-2)^{-1}R_{\mu\nu}k^\mu k^\nu$. 
It follows that the divergence of the Ricci tensor component $R_{\mu\nu}k^\mu k^\nu$
implies the p.p curvature singularity \cite{paperI}. For later sections, we repeatedly encounter this kind of singularity. 

The causal structure of the spacetime is essentially encoded into the two-dimensional 
part $\D s_2^2=-f_1(r)\D t^2+f_2(r)\D r^2$ of the metric, which is recast into 
\begin{align}
\label{tortoise}
\D s_2^2=-f_1(r)\D t^2+f_2(r)\D r^2
=-f_1(r_*(r))(\D t^2-\D r_*^2)\,, \qquad r_*=\int^r \sqrt{\frac{f_2(r)}{f_1(r)}}\D r \,. 
\end{align}
Since this coordinate manifests the conformal flatness, one can 
extract the causal structure just by a comparison with that in Minkowski spacetime. 
In short, if the tortoise coordinate $r_*$ blows up at some locus $r$, 
it corresponds to the null surface. If $r_*$ fails to diverge, 
it corresponds to the timelike (spacelike) surface for $f_1(r_*(r))>0$ ($f_1(r_*(r))<0$). 

In what follows, we enumerate three solutions belonging to the class (\ref{metric}) 
and discuss the fundamental aspects.

\subsubsection{Fisher solution}

The $n(\ge 4)$-dimensional Fisher class solution \cite{Fisher:1948yn} reads  
\begin{subequations}
\label{FJNW}
\begin{align}
\D s^2=&-f(r)^{\alpha}\D \tau^2+ f(r)^{-(\alpha+n-4)/(n-3)}\biggl(\D r^2+r^2f(r)\D \Omega_{n-2}^2 \biggl),\\
\phi=&\pm \sqrt{{\epsilon}\frac{(n-2)(1-\alpha^2)}{4(n-3)\kappa_n}}\ln f(r),\qquad f(r)=1-\frac{M}{r^{n-3}}\,, 
\end{align}
\end{subequations}
where $\alpha$ and $M$ are constants. We have chosen 
the asymptotic value of the scalar field to vanish for definiteness of the argument, 
which can be easily restored by $\phi\to \phi-\phi_0$. 
This solution in four dimensions reduces to the original  solution found by Fisher~\cite{Fisher:1948yn}. The higher dimensional generalization can be found in~\cite{JNWhigher}. 
For the ordinary (phantom) scalar case $\epsilon=1$ ($\epsilon=-1$), 
we have $\alpha^2 \le 1$ ($\alpha^2 > 1$). By inspecting the form of the metric, 
one finds that the coordinate change $r^{n-3}\to r^{n-3}+M$
leads to the following reflection symmetry
\begin{align}
\label{Malpha}
M\to -M \,, \qquad \alpha \to -\alpha\,, 
\end{align}
which will be used to focus on the $M>0$ case. In due course, we will see that this symmetry 
carries over to the asymptotically (A)dS solutions. 
The combination  $\alpha M$ only contributes to the Arnowitt-Deser-Misner (ADM) mass \cite{Arnowitt:1962hi}
$M_{\rm ADM}\propto \alpha M$.

When $\alpha^2=1$, the Fisher solution (\ref{FJNW}) reduces to the Schwarzschild-Tangherlini vacuum solution.  As shown in \cite{paperI}, the Fisher class solution with $\alpha^2\ne 1$ is always singular, in that it inescapably admits a scalar curvature singularity or a p.p curvature singularity at $r=M^{1/(n-3)}$ (for $M>0$) which is not covered by a horizon.

For later purposes, let us rewrite the Fisher solution (\ref{FJNW}) in the isotropic coordinate system as
\begin{subequations}
\label{JNWiso}
\begin{align}
\D s^2&= - \left(\frac{1-\frac{M}{4\rho^{n-3}}}{1+\frac{M}{4\rho^{n-3}}}\right)^{2\alpha} \D \tau^2
+\left(1+\frac{M}{4\rho^{n-3}}\right)^{\frac{4}{n-3}} 
\left(\frac{1-\frac{M}{4\rho^{n-3}}}{1+\frac{M}{4\rho^{n-3}}}\right)^{\frac{2(1-\alpha)}{n-3}} 
(\D \rho^2+\rho^2 \D \Omega_{n-2}^2) \,, \\
\phi &=\pm \sqrt{{\epsilon}\frac{(n-2)(1-\alpha^2)}{(n-3)\kappa_n}}\ln \left|\frac{1-\frac{M}{4\rho^{n-3}}}{1+\frac{M}{4\rho^{n-3}}}\right|\,, 
\end{align}
\end{subequations}
where 
$\rho $ is defined by the relation
\begin{align}
\label{}
r=\rho \left(1+\frac{M}{4\rho^{n-3}}\right)^{{2}/(n-3)} \,.
\end{align}

\subsubsection{Ellis-Gibbons solution}

The next class of solutions we consider is the $n(\ge 4)$-dimensional Ellis-Gibbons solution~\cite{Ellis1973,Gibbons:2003yj}. 
This is a solution for the phantom case ($\epsilon=-1$) only and is given by 
\begin{align}
\label{EllisGibbons}
\D s^2=&-e^{-M/r^{n-3}}\D \tau^2+e^{M/[(n-3)r^{n-3}]}\bigl(\D r^2+r^2\D \Omega_{n-2}^2\bigl)\,,\qquad 
\phi=\pm \sqrt{-{\epsilon}\frac{n-2}{4(n-3)\kappa_n}}\frac{M}{r^{n-3}}\,, 
\end{align}
where $M$ being a constant proportional to the ADM mass.

As pointed out by Gibbons \cite{Gibbons:2003yj} in four dimensions, the 
above metric can be generalized to have multiple point sources as 
\begin{align}
\D s^2=-e^{-H}\D \tau^2+e^{H/(n-3)}h_{IJ}\D y^I\D y^J,\qquad 
\phi=\pm \sqrt{-{\epsilon}\frac{n-2}{4(n-3)\kappa_n}}H\,, 
\label{Gibbons}
\end{align}
where $h_{IJ}$ is an arbitrary $(n-1)$-dimensional Ricci-flat Riemannian metric ${}^{(n-1)}{R}_{IJ}=0$ and $H$ is a harmonic function $\Delta_h H=0$. If we consider the flat space $h_{IJ}\D y^I \D y^J=\D r^2+r^2 \D \Omega_{n-2}^2$ and restrict only to the spherical harmonics, the Gibbons solution (\ref{Gibbons}) goes back to the Ellis-Gibbons solution~(\ref{EllisGibbons}).

As demonstrated in a companion paper \cite{paperI},  the Ellis-Gibbons solution (\ref{EllisGibbons})
is always singular at $r=0$, regardless of the sign on $M$ and the spacetime dimensionality. 
The divergence of the scalar curvature only occurs for $M<0$. In the positive mass case ($M>0$), there appears alternatively a naked p.p curvature singularity at $r=0$. Moreover, the $r=0$ surface for $M>0$ can be reachable within a finite affine time for the radial null geodesics. This means that the $r=0$ surface is not null infinity, despite the divergence of the areal radius.  
These properties thereby rule out the possibility that the Eills-Gibbons solution (\ref{EllisGibbons}) describes a regular wormhole spacetime 
(see the argumentation e.g., in \cite{Boonserm:2018orb}).  
As we will discuss later, this situation changes considerably, provided the scalar potential is introduced, 
for which the singularity at $r=0$ may be covered by the event horizon of a black hole.

\subsubsection{Ellis-Bronnikov solution}

The third class of spacetimes is the $n(\ge 4)$-dimensional Ellis-Bronnikov solution~\cite{Ellis1973,Bronnikov1973,paperI}. 
This solution exists only in the phantom case ($\epsilon=-1$)  and is given by
\begin{subequations}
\label{Ellis-Bronnikov}
\begin{align}
\D s^2&=-e^{-2\beta U(r)}\D \tau^2+e^{2\beta U(r)/(n-3)}V(r)^{1/(n-3)}\biggl(\frac{\D r^2}{V(r)} +r^2
\D \Omega_{n-2}^2 \biggl),\\
\phi&=\pm\sqrt{-{\epsilon}\frac{(n-2)(1+\beta^2)}{\kappa_n(n-3)}} U(r) \,,
\end{align}
\end{subequations}
where 
\begin{align}
\label{UV}
U(r) \equiv \arctan \left(\frac{M}{2r^{n-3}}\right) \,, 
 \qquad V(r)\equiv  1+\frac{M^2}{4r^{2(n-3)}}\,.
\end{align}
The solutions is specified by two parameters $\beta$ and $M$.
By virtue of the symmetry
\begin{align}
\label{Mbeta}
M\to -M \,, \qquad \beta \to -\beta \,, 
\end{align}
one can confine to the $M>0$ case to explore the 
physical properties of the solution. The ADM mass is controlled by 
a combination $M_{\rm ADM}\propto \beta M$.

As explained in our previous paper \cite{paperI}, the Ellis-Bronnikov solution is entitled to be a 
traversable wormhole which bridges the two asymptotically flat regions.
The $r=0$ surface is merely a coordinate singularity. The extension through 
$r=0$ is best achieved by the replacement $U(r)\to\pi/2-\arctan(2r^{n-3}/M)$ and $x=r^{n-3}$.\footnote{
Remark $\arctan(x)+\arctan(1/x)=\pi/2 $ for $x>0$.}
 The resulting metric is smooth at $x=0$ and can be extended into the 
$x<0$ region. This yields a maximal extension of the spacetime with 
two asymptotically flat regions, i.e., the spacetime describes a regular wormhole. 
It deserves to remark that this extension is asymmetric for $\beta \ne 0$, 
since the ADM mass in each asymptotic region has unequal value.  
The symmetric extension only occurs for the zero mass wormhole $\beta =0$.

In terms of the isotropic coordinate $\rho$ defined by 
\begin{align}
\label{}
r=\rho \left(1-\frac{M^2}{16\rho^{2(n-3)}}\right)^{1/(n-3)} \,, 
\end{align}
the {Ellis-Bronnikov} solution (\ref{Ellis-Bronnikov}) can be transformed into
\begin{subequations}
\label{Ellis-Bronnikov:iso}
\begin{align}
\D s^2&=- e^{-2\beta \hat U(\rho)}\D \tau^2+ e^{2\beta \hat U(\rho)/(n-3)}\left(
1+\frac{M^2}{16\rho^{2(n-3)}} 
\right)^{2/(n-3)}(\D \rho^2+\rho^2 \D \Omega_{n-2}^2)\,, \\
\phi&=\pm\sqrt{-\epsilon\frac{(n-2)(1+\beta^2)}{\kappa_n(n-3)}} \hat U(\rho)\,, 
\end{align}
\end{subequations}
where $\hat U(\rho)=U(r(\rho))$. 
This form of the metric will be of use in obtaining asymptotically (A)dS solutions.

\subsection{Toward (A)dS generalization}
\label{sec:(A)dS}

In the previous subsection, we have briefly reviewed static and spherical solutions with a 
massless phantom scalar field. In the subsequent sections, we explore their generalizations in (A)dS spacetimes. 
For the nontrivial (A)dS asymptotics, we need the scalar field potential. Consequently, 
we are concerned with the system described by the action\footnote{We apologize the reader that we use the same letter $V$ for the potential as the one in (\ref{UV}). Nonetheless, we expect that no confusion arises.}
\begin{align}
\label{action:pot}
S=\int \D ^nx \sqrt{-g} \left(\frac{R}{2\kappa_n}
-\frac 12 \epsilon(\nabla \phi)^2 -V(\phi) 
\right)\,,
\end{align}
As discussed at the introduction, no systematic process is available in the construction of solutions with a scalar potential or a cosmological constant. 
Nevertheless, our proposed procedure works properly and is compiled as follows:
\begin{itemize}
  \item[(i)] Write the static seed metric in Einstein-phantom scalar system by means of the isotropic coordinates.
  \item[(ii)] Insert the scale factor $a(\tau)=e^{H\tau}$ into the metic and the scalar field configuration 
  in such a way that the metric asymptotes to the 
dS universe with a Hubble parameter $H$ in a flat chart, and that the solution is invariant under the generalized time translation (see (\ref{taurho}) below). 
  \item[(iii)] Substitute the ansatz into field equations derived by the action (\ref{action:pot}) and 
  determine the scalar field potential $V=V(\phi)$. 
  \item[(iv)] Exploit the newly appeared symmetry in the metric ansatz at step (ii) to bring the solution into the manifestly static, asymptotically dS form.
 \item[(v)] Perform the Wick rotation $H\to -i \ell^{-1}$ of the Hubble parameter to obtain the 
 asymptotically AdS solutions. 
\end{itemize}

Adapting this guiding principle to the Schwarzschild solution, one can derive the 
Schwarzschild-(A)dS solution, which has a constant scalar potential, i.e., the scalar field is trivial and the potential 
is nothing but the cosmological constant. 
When the seed metric solves Einstein's equations with a massless (phantom-)scalar field, 
the new solution derived in the above prescription solves Einstein's equations sourced by 
a  (phantom-)scalar field with a potential. 
To precisely compensate the expansion of the universe by the 
introduction of the potential of the scalar field, it is essential to write the metric 
in terms of the isotropic coordinate (step (i)). It seems that other gauges of radial coordinate do not work. 

Since the potential of the scalar field is derived in the above prescription, 
one may envisage that it gives rise to unusual form, which has no 
origin in fundamental physics. 
An intriguing outcome here is that  all the potentials $V(\phi)$ obtained in the above procedure  are represented in terms of the ``superpotential'' $W(\phi)$ as 
\begin{align}
\label{pot:superpotential}
V(\phi)=2(n-2)\left(\frac{\epsilon}{\kappa_n}(n-2)\left(\frac{\partial}{\partial \phi}W(\phi)\right)^2-(n-1)W(\phi)^2 \right)\,. 
\end{align}
For $\epsilon=+1$ and real $W(\phi)$, this form of the potential commonly appears in supergravity and in the positive energy theorems \cite{Boucher:1984yx,Townsend:1984iu,Nozawa:2013maa,Nozawa:2014zia}. 
The critical point of the superpotential is the extremum of the potential, but the converse is not true. 
As we will see later, the function $W(\phi)$ turns out purely imaginary for the solutions asymptotic to dS/FLRW universes. In contrast, the superpotential can be made to be real 
for the asymptotically AdS case. 

We also note that the final expression of the static metric in step (iv) and (v) can be further 
used to put an ansatz to generate yet another new solution. In the subsequent sections, we determine the potential for each class of solutions and discuss physical/causal properties. 

%
%Starting from the solutions obtained in the body of text, we can generate 
%new solutions with different asymptotic structure, like (anti-)de Sitter or 
%FLRW universe. This construction is not exhaustive, but nevertheless 
%this generation procedure is of interest in an attempt to make a complete catalogue solutions. 

%Here we shall not attempt to discuss the physical properties of these solutions, which will be discussed in a forthcoming paper. 

\section{Fisher class of solutions in (A)dS}
\label{sec:Fisher}

We begin by the analysis of the Fisher class of solutions. The whole technical scheme expanded here can be straightforwardly used also in the subsequent sections.

\subsection{Fisher solution in dS: plus branch}

We suppose the following ansatz for the generalization of the Fisher solution (\ref{FJNW})
into dS space 
\begin{subequations}
\label{JNWdyn}
\begin{align}
\D s^2=& - \left(\frac{1-\frac{M}{4(a(\tau)\rho)^{n-3}}}{1+\frac{M}{4(a(\tau)\rho)^{n-3}}}\right)^{2\alpha} \D \tau^2
\notag \\
&+a^2(\tau)\left(1+\frac{M}{4(a(\tau)\rho)^{n-3}}\right)^{\frac{4}{n-3}} 
\left(\frac{1-\frac{M}{4(a(\tau)\rho)^{n-3}}}{1+\frac{M}{4(a(\tau)\rho)^{n-3}}}\right)^{\frac{2(1-\alpha)}{n-3}} 
(\D \rho^2+\rho^2 \D \Omega_{n-2}^2) \,, \\
\phi = &+\sqrt{{\epsilon}\frac{(n-2)(1-\alpha^2)}{(n-3)\kappa_n}}
\ln \left|\frac{1-\frac{M}{4(a(\tau)\rho)^{n-3}}}{1+\frac{M}{4(a(\tau)\rho)^{n-3}}}\right|\,, 
\end{align}
\end{subequations}
where $a(\tau)\equiv e^{H\tau}$ and $H$ is a constant playing the role of Hubble parameter. 
When $H=0$, the above solution reduces to the 
spherical Fisher solution written in the isotropic coordinate (\ref{JNWiso}). 
Here we have chosen the ``plus sign'' of the scalar field for the sake of expedience. We shall therefore call the solution 
(\ref{JNWdyn}) as the ``plus branch.''
As $\rho\to \infty$, the solution (\ref{JNWdyn}) approaches to the dS universe in the flat spatial patch. 
This metric reduces to the Schwarzschild-dS solution in the McVittie's form \cite{McVittie:1933zz}
for $\alpha^2=1$.

Plugging this ansatz into the gravitational field equations derived from the action (\ref{action:pot}), 
the potential for the scalar field is determined to be 
\begin{align}
\label{potential:JNW}
V(\phi)=&\frac{n-2}{4\kappa_n}H^2 e^{-2\alpha \alpha_0 \phi} \Bigl[
2(n-2)(1-\alpha^2)+[(n-2)\alpha -1](\alpha-1)e^{-2\alpha_0 \phi}\notag \\
&
+[(n-2)\alpha +1](\alpha+1)e^{2\alpha_0\phi} 
\Bigr]\,, 
\end{align}
where 
\begin{align}
\label{}
\alpha_0\equiv \sqrt{\frac{(n-3)\kappa_n}{(n-2)\epsilon(1-\alpha^2)}} \,. 
\end{align} 
Notably, this potential is built out of the following  superpotential
\begin{align}
\label{JNW:superpotential}
W(\phi)=\frac{i H}{2\sqrt{\kappa_n}} e^{-\alpha \alpha_0 \phi} 
\Bigl(\cosh (\alpha_0\phi)+\alpha \sinh (\alpha_0\phi)\Bigr)\,,
\end{align}
as (\ref{pot:superpotential}). 
Let us underline that the parameter $\alpha$ is no longer a parameter of the solution, 
as opposed to the asymptotically flat case (\ref{FJNW}).
Rather, it is promoted to be a parameter to define the theory (\ref{potential:JNW}).

We can verify that the origin  for the scalar field $\phi=0$ is the critical point of the potential corresponding to the dS vacuum with the Hubble parameter $H$, as well as the critical point of the superpotential. 
The potential may admit another critical point at 
$\phi=2\sqrt{(n-2)(1-\alpha^2)\epsilon/[(n-3)\kappa_n]}\ln[((n-2)\alpha-1)/((n-2)\alpha+1)]$ when inside the logarithm is positive. 
However, this  does not play a role in our current analysis.

The solution (\ref{JNWdyn}) can be viewed as an interpolating solution between 
the Fisher solution (\ref{FJNW}) and the dS universe. At first glance, one may expect that this 
is a dynamical solution. 
Nevertheless,  the solution (\ref{JNWdyn}) is invariant under
\begin{align}
\label{taurho}
\tau \to \tau +c\,,\qquad \rho\to \rho e^{-Hc}\,, \qquad 
\textrm{($c$: constant)}\,,
\end{align} 
leading to  the existence of a  
Killing vector $\xi^\mu=(\partial/\partial \tau )^\mu-H \rho (\partial/\partial \rho)^\mu$. 
One can then bring the metric into a manifestly static form
\begin{align}
\label{}
\D s^2 &= -f_{\rm F}(\hat \rho)\frac{\hat f_-(\hat\rho)^{2\alpha}}{\hat f_+(\hat\rho)^{2\alpha}}\D t^2+\hat f_+(\hat \rho)^{2(1+\alpha)/(n-3)}
\hat f_-(\hat \rho)^{2(1-\alpha)/(n-3)}\left(\frac{\D \hat \rho^2}{f_{\rm F}(\hat \rho)}
+\hat\rho^2 \D \Omega_{ n-2}^2 
\right)\,, \\
\phi &= \sqrt{{\epsilon}\frac{(n-2)(1-\alpha^2)}{(n-3)\kappa_n}}
\ln \left|\frac{\hat f_-(\hat \rho)}{\hat f_+(\hat \rho)}\right|\,, 
\end{align}
by the coordinate transformations
\begin{align}
\label{}
 t=\tau  +\int \frac{H \hat \rho \hat f_+(\hat \rho)^{2[1+(n-2)\alpha]/(n-3)}
\hat f_-(\hat \rho)^{2[1-(n-2)\alpha]/(n-3)}}{f_{\rm F}(\hat \rho)}\D \hat \rho\,, \qquad 
\hat \rho= \rho e^{H\tau } \,, 
\end{align}
with 
\begin{align}
\label{}
f_{\rm F}(\hat \rho) &\equiv  
1-H^2 \hat \rho^2 
\hat f_+(\hat \rho)^{2[1+(n-2)\alpha]/(n-3)}
\hat f_-(\hat \rho)^{2[1-(n-2)\alpha]/(n-3)}\,, 
\\
\hat f_\pm (\hat \rho)&\equiv 1\pm \frac{M}{4\hat \rho^{n-3}} \,. 
\end{align}
Performing the further coordinate transformation 
\begin{align}
\label{}
\hat \rho^{n-3}=\frac{M}{4} \frac{1+\sqrt{1-M/r^{n-3}}}{1-\sqrt{1-M/r^{n-3}}} \,, 
\end{align}
the solution can be brought into a more familiar form 
\begin{subequations}
\label{JNWdS:static}
\begin{align}
\D s^2=&-f(r)^{\alpha}F_{\rm F}^+(r) \D t^2 +f(r)^{-(\alpha+n-4)/(n-3)}\left(
\frac{\D r^2}{F_{\rm F}^+(r)}+r^2 f(r) \D \Omega_{ n-2}^2\right) \,, \\
\phi=&  \sqrt{{\epsilon}\frac{(n-2)(1-\alpha^2)}{4(n-3)\kappa_n}} \ln f(r)\,, 
\end{align}
\end{subequations}
where $f(r)= 1-M/r^{n-3}$ as before and 
\begin{align}
\label{JNWdS:staticFH}
F_{\rm F}^+(r)\equiv  1- H^2 r^2f(r)^{-[(n-2)\alpha-1]/(n-3)}\,.
\end{align}
The $\alpha^2=1$ case reduces to the Schwarzschild-dS solution in the static patch. 
This form of  the metric convinces one to recognize that this is a natural generalization of Fisher solution in dS space. We explore below the causal and physical properties of the solution (\ref{JNWdS:static}). For definiteness of the argument, we suppose $H>0$ henceforth.

We first remark that the metric (\ref{JNWdS:static}) is unaltered under 
the simultaneous sign flip (\ref{Malpha}) of ($M, \alpha$). 
Thanks to this symmetry, we can focus on the $M>0$ case, for which 
 the interior boundary of $r$ is $r_s \equiv M^{1/(n-3)}$.

The Ricci scalar of the solution (\ref{JNWdS:static}) reads
\begin{align}
\label{RicciJNWdS}
R=&\frac {M^2}{4r^{2(n-2)}}(n-2)(n-3)(1-\alpha^2)f^{(\alpha-n+2)/(n-3)}\notag \\
&+\frac 14 H^2 (n-1)
\left[4n-\frac{4nM}{r^{n-3}}(1+\alpha)+\frac{M^2}{r^{2(n-3)}}(1+\alpha)
\bigl(6-6\alpha+n(3\alpha-1)\bigr)\right] f^{-1-\alpha} \,. 
\end{align}
One finds that the $r=r_s$ surface 
corresponds to the curvature singularity for any parameter region of $\alpha (\ne \pm 1)$. 
This is in sharp contrast to the $H=0$ case, for which 
the $r=r_s$ surface might correspond merely to a p.p curvature singularity for some parameter region, 
rather than the scalar curvature singularity. 

Let us next investigate the causal properties of this singularity. 
Since the function $F^+_{\rm F}(r)$ does not contribute to the equation (\ref{affine}) that determines the affine parameter of the radial null geodesics, one can use the criterion for the $H=0$ case if the surface $r=r_s$ resides at an infinite affine distance for the radial null geodesics. Inferring from the results in~\cite{paperI}, $r=r_s$ is null infinity for $\alpha \le -(n-2)/(n-4)$ for $n\ge 5$, otherwise it locates at finite affine distance. 
For the signature of the surface $r=r_s$, eq. (\ref{tortoise}) gives rise to 
$r_*=\int [f(r)^p/F_{\rm F}^+(r)]\D r$ with $p=-[(n-2)\alpha+n-4]/[2(n-3)]$, leading to 
three cases to consider (i) $\alpha<1/(n-2)$, (ii) $\alpha =1/(n-2)$ and (iii) $\alpha>1/(n-2)$. 
In case (i), $F_{\rm F}^+(r_s)=1$, so that $r_*$ remains finite as $r\to r_s$, implying  that 
$r=r_s$ is timelike. In case (ii), $r=r_s$ is timelike (spacelike) for $H<1/r_s$ ($H>1/r_s$). 
In case (iii), one finds that $r=r_s$ is spacelike.

Another notable feature in the $H\ne 0$ case is that  the solution (\ref{RicciJNWdS}) may admit a black hole horizon $r=r_+$ at $F_{\rm F}^+(r_+)=0$,  on top of the cosmological horizon $r=r_c(>r_+)$. One can check that any curvature invariants remain finite at the surfaces $F_{\rm F}^+(r)=0$, i.e, these are 
regular null hypersurfaces and constitute Killing horizons for the Killing vector $\partial/\partial t$. 
The condition for the presence of the event horizon  boils down to 
\begin{align}
\label{JNWdS:horizoncond1}
\alpha> \frac{1}{n-2} \,, \qquad 
0<M < \frac{\alpha_>^{\alpha_>}}{H^{n-3}(1+\alpha_>)^{1+\alpha_>}}\,,
\end{align}
where $\alpha_>\equiv [(n-2)\alpha-1]/2>0$. 
The first condition ensures that $r_+>r_s\equiv M^{1/(n-3)}$, whereas the second condition 
implies $r_c>r_+$. 
It follows that the event horizon exists both for phantom ($\alpha^2>1$) and 
non-phantom ($\alpha^2 <1$) cases.

If the parameters fail to fulfill the criteria above, we have two conceivable cases. When the first condition in 
(\ref{JNWdS:horizoncond1}) is false, the equation $F_{\rm F}^+(r)=0$ admits only a single root, corresponding to the cosmological horizon,  outside the singularity $r=r_s$. This class of spacetime is static in the $r<r_c$ region and describes a naked singularity. The other case is that the second condition in
(\ref{JNWdS:horizoncond1})  is false, while the first condition is met. In this case, $F_{\rm F}^+(r)$ is negative-definite, meaning that the spacetime 
is not static in the sense that the Killing vector $\partial/\partial t$ is spacelike anywhere. 
This class of spacetime looks like the interior of the Schwarzschild black hole and exhibits the dynamical aspect.  

\begin{figure}[t]
\begin{center}
\includegraphics[width=7cm]{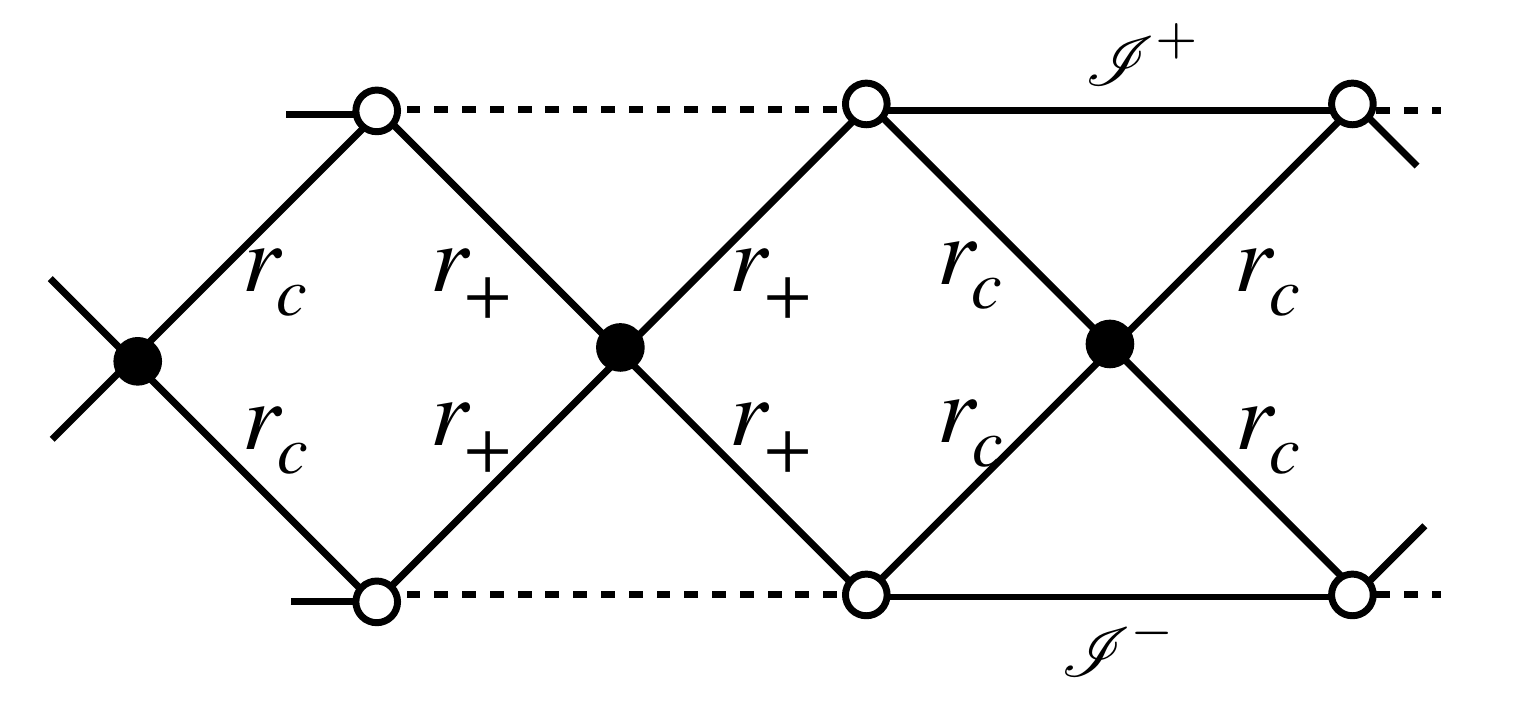}
\caption{Possible conformal diagram for the dS Fisher solutions (plus (\ref{JNWdS:static}) and minus (\ref{JNWdS:static2}) branches) and the 
dS Ellis-Gibbons solutions (plus (\ref{GibbonsdSstatic1}) and minus (\ref{GibbonsdSstatic2}) branches) admitting the event horizon of a  black hole. 
The required parameter region is (\ref{JNWdS:horizoncond1}), (\ref{FdSminus:horizon}),  (\ref{GibbonsdSplus:horizon})
(\ref{Gibbonsmassrange2}), respectively. 
$\mas I^\pm$ denote future and past null infinities and dashed lines correspond to the 
scalar curvature singularities. 
White circles stand  for timelike infinities, while the black circles stand 
for the bifurcation surfaces, respectively.
}
\label{fig:PDFdS}
\end{center}
\end{figure}

We present in figure \ref{fig:PDFdS} the Penrose diagram for the plus-branch Fisher solution in dS universe satisfying (\ref{JNWdS:horizoncond1}) which admits a black hole horizon.
The global structure is identical to the Schwarzschild-dS black hole. 
Throughout the paper, we shall not display the causal diagrams for singular configurations. 

\subsection{Fisher solution in AdS: plus branch}

Since we have a static form of the metric (\ref{JNWdS:static}) at hand, one can Wick rotate the Hubble parameter $H \to -i \ell^{-1}$ to give an 
asymptotically AdS solution, i.e., 
$F_{\rm F}^+(r)$ in (\ref{JNWdS:staticFH}) is now given by
\begin{align}
\label{FFp}
F_{\rm F}^+(r)\equiv  1+\frac{ r^2}{\ell^2 }f(r)^{-[(n-2)\alpha-1]/(n-3)}\,. 
\end{align}
The solution  (\ref{JNWdS:static}) with (\ref{FFp}) solves the field equations derived from the action
(\ref{action:pot}), where the potential is given by (\ref{potential:JNW}) with $H\to -i\ell^{-1}$. 
The parameter $\ell$ has a dimension of the  length  and corresponds to the AdS radius. 
 This solution is  contained, for $\epsilon=1$,  in the class of metrics with topological base found in \cite{Feng:2013tza} (set $H_{1\ \rm there}=f(r)$, $H_{2 \ \rm there}=1$ and 
$\alpha_{\rm there}=0$ in (20) in \cite{Feng:2013tza}). 
For the spherical case, the corresponding asymptotically AdS solution never admits an event horizon of a black hole for either sign of $\epsilon$,  by virtue of $F_{\rm F}^+(r)>0$.  
Thus, the solution describes a naked singularity at $r=M^{1/(n-3)}$.

\subsection{Fisher solution in dS: minus branch}

Following the observation in \cite{Faedo:2015jqa}, 
one can obtain yet another new solution in the theory (\ref{potential:JNW}).
We reverse the sign of $\phi$ in (\ref{JNWdS:static}), consider the metric 
(\ref{JNWdS:static}) with $f(r)=1-M/r^{n-3}$ and leave 
$F_{\rm F}^+(r)$ unspecified. Since the potential is now given by  (\ref{potential:JNW}), 
one can determine this unknown function $F_{\rm F}^+(r)$ from Einstein's equations. 
It turns out that this procedure works only in $n=4$,  yielding 
\begin{subequations}
\label{JNWdS:static2}
\begin{align}
\D s^2=&-f(r)^{\alpha}F_{\rm F}^-(r) \D t^2 +f(r)^{-\alpha}\left(
\frac{\D r^2}{F_{\rm F}^-(r)}+r^2 f(r) \D \Omega_{n-2}^2\right) \,, \\
\phi=& - \sqrt{{\epsilon}\frac{1-\alpha^2}{2\kappa_4}} \ln f(r)\,, 
\end{align}
\end{subequations}
where $f(r)=1-M/r$ and 
\begin{align}
\label{Fminus}
F_{\rm F}^-(r) \equiv 
1-H^2\Bigl(r^2-M(1-2\alpha)r-M^2\alpha(1-2\alpha)\Bigr)\,.
\end{align}
From the minus sign of the scalar field in (\ref{JNWdS:static2}), 
we refer to the solution as the ``minus branch.'' 
In the massless scalar case, this sign choice was the matter of convention. 
In the presence of the scalar field potential, this sign is crucial and each solution 
describes the different spacetime. 

The $\alpha^2=1$ reduces to the Schwarzschild-dS solution, while $M=0$ describes the 
dS metric.  
The solution (\ref{JNWdS:static2}) also enjoys the reflection symmetry $\alpha \to -\alpha$ with 
$M\to -M$, which allows us to set $M>0$. At the surface $r=M(>0)$,
the curvature invariants diverge for $\alpha < 2$, while it corresponds to the 
p.p. curvature singularity for any $\alpha(\ne \pm 1)$. This can be deduced by 
\begin{align}
\label{RFdSminus}
R=&\frac{M^2(1-\alpha^2)}{2r^4}f(r)^{\alpha-2}+\frac{H^2 }{2r^4}f(r)^{\alpha-2}\Bigl[
24r^3\left(r+(\alpha-2)M\right)\notag \\
&+9M^2 r^2(\alpha-3)(\alpha-1)+M^3 r(\alpha-3)(\alpha-1)(2\alpha-1)
+M^4\alpha (\alpha^2-1)(2\alpha-1)
\Bigr]\,, 
\end{align}
and
\begin{align}
\label{RkkFdSminus}
R_{\mu\nu}k^\mu k^\nu =-\frac{M^2(\alpha^2-1)}{2(r-M)^2r^2}\,, 
\end{align}
where $k^\mu =(f F_{\rm F}^-)^{-1}(\partial/\partial t)^\mu +(\partial/\partial r)^\mu$
is the affine-parametrized radial geodesic tangent. 
It is interesting to observe that the singular nature of $f(r)=0$ surface is 
substantially different from the plus-branch solution, despite the apparent similarity of the metric. 
In the entire parameter region, the minus-branch solution is always singular at
$r=M(>0)$, even though the curvature invariants remain finite there for $\alpha\ge 2$.

Around $r=M$, the tortoise coordinate is approximated by
(we exclude the $F_{\rm F}^-(M)=0$ case since $r=M$ is singular)
\begin{align}
\label{}
r_*=\int ^r \frac{\D r}{f(r)^\alpha F_{\rm F}^-(r)} \simeq \frac{1}{f'(M)^\alpha F_{\rm F}^-(M)} \int 
(r-M)^{-\alpha} \D r \,.
\end{align}
It follows that the $r=M$ is null for $\alpha \ge 1$. 
Since $F_{\rm F}^-(M)=1-H^2M^2 \alpha(1+2\alpha)$, 
the $r=M$ surface is spacelike for 
\begin{align}
\label{FdSminus:signature}
0<\frac{1}{H\sqrt{\alpha(1+2\alpha)}}<M\,.
\end{align}
Obviously this is valid for $\alpha<-1/2$ or $0<\alpha(<1)$. 
In other parameter range, the $r=M$ surface is timelike.

Now we turn our attention to the horizon at $F_{\rm F}^-(r)=0$. 
The condition for the appearance of the event horizon at $r=r_+(>M>0)$ boils down to
\begin{align}
\label{FdSminus:horizon}
\alpha <-\frac 12  \,, \qquad 
0<\frac{1}{H\sqrt{\alpha(1+2\alpha)}}<M<\frac{2}{H\sqrt{4\alpha^2-1}} \,,
\end{align}
which assures $0<M<r_+<r_c$, where $r_+$ and $r_c$ are respectively the loci of the event and the cosmological horizon and are give by 
\begin{align}
\label{}
r_+=\frac M2 \left(1-2\alpha-\sqrt{4(MH)^{-2}+1-4\alpha^2}\right)\,, \qquad 
r_c=\frac M2 \left(1-2\alpha+\sqrt{4(MH)^{-2}+1-4\alpha^2}\right)\,,
\end{align}
which satisfy $F_{\rm F}^-(r_{+})=F_{\rm F}^-(r_{c})=0$.
The event horizon exists both for the phantom and non-phantom cases. 
Inspecting (\ref{FdSminus:signature}), the singularity is spacelike in the parameter range 
(\ref{FdSminus:horizon}). The Penrose diagram is therefore the same as in figure \ref{fig:PDFdS}.

\subsection{Fisher solution in AdS: minus branch}

Upon replacement $H\to -i \ell^{-1}$ for $\epsilon=+1$, 
the solution (\ref{JNWdS:static2}) recovers the $\mathcal N=2$ supergravity solution in 
AdS, which allows a parameter range under which the solution possesses 
the event horizon~\cite{Faedo:2015jqa}. 
The function $F^-_{\rm F}(r)$ in (\ref{Fminus}) is now 
\begin{align}
\label{minusFisher}
F_{\rm F}^-(r) =
1+\frac 1{\ell^2}\Bigl(r^2-M(1-2\alpha)r-M^2\alpha(1-2\alpha)\Bigr)\,.
\end{align}
We suppose $\ell>0$ (and $M>0$) in the hereafter. 
The conditions under which the non-phantom solution ($\epsilon=1$ and $\alpha^2\le 1$) enjoys the event horizon of a black hole were already addressed in~\cite{Faedo:2015jqa}. Here, we discuss this issue in 
wider range of parameters.

As in the dS case, the $r=M(>0)$ surface is singular in that it corresponds to the scalar curvature singularity ($\alpha<2$) or the p.p curvature singularity (see (\ref{RFdSminus}) and (\ref{RkkFdSminus})). 
The radial coordinate $r$ is identified as an affine parameter for the radial null geodesics. 
Deducing from $r_*=\int \D r/(f(r)^\alpha F_{\rm F}^-(r))$, 
the $r=M(>0)$ surface is null for $\alpha \ge 1$, while it is spacelike 
for $M>\ell/\sqrt{-\alpha(1+2\alpha)}$ (requiring $-1/2<\alpha<0$) and timelike otherwise.

\begin{figure}[t]
\begin{center}
\includegraphics[width=4.5cm]{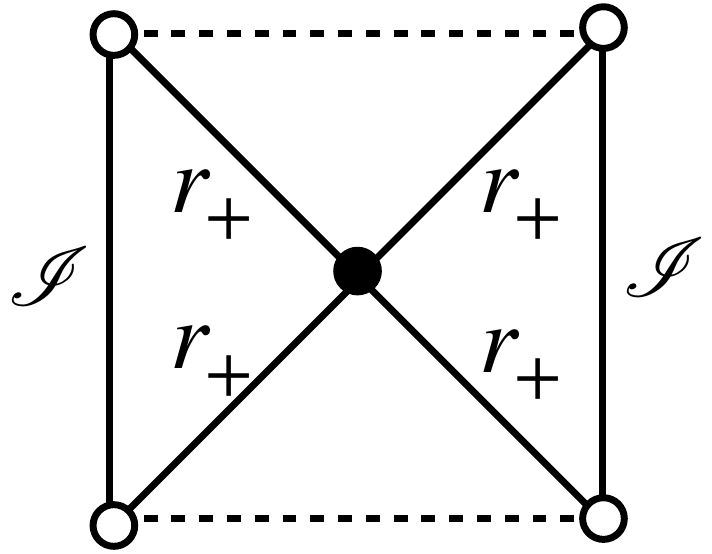}
\caption{A conformal diagram for the minus-branch AdS Fisher solution (\ref{minusFisher}) admitting the event horizon of a  black hole. The condition for the existence of the horizon is (\ref{FAdSminus:horizon}), under which the central singularity at $r=M$ is spacelike. $\mas I$ in the figure represents the AdS infinity. 
}
\label{fig:PDFAdS}
\end{center}
\end{figure}

We have horizons at $r=r_\pm$, where 
\begin{align}
\label{}
r_\pm= \frac {1}{2} \left((1-2\alpha)M\pm \sqrt{(1-4\alpha^2)M^2-4\ell^2}\right)\,. 
\end{align}
For the solution to represent a black hole which is regular on and outside the event horizon $r=r_+$, 
we must have $r_+>M(>0)$. This condition reduces to 
\begin{align}
\label{FAdSminus:horizon}
M>\frac{\ell}{\sqrt{-\alpha (1+2\alpha)}} >0\,.
\end{align}
This demands $-1/2<\alpha <0$, i.e, 
the phantom case results in the spacetime with a naked singularity. 
Since the scalar field is non-phantom for the existence of the event horizon,  we can refer to \cite{Faedo:2015jqa} for 
the physical property of the black hole solution. 
In the parameter region (\ref{FAdSminus:horizon}) we have 
$M>r_-$, so that the singularity at $r=M$ is spacelike. 
It turns out that the global structure of the black hole solution is the 
same as the Schwarzschild-AdS black hole, see figure \ref{fig:PDFAdS}.

\section{Gibbons class of solutions in (A)dS}
\label{sec:Gibbons}

We shall next consider the (A)dS generalization of the Gibbons class (\ref{Gibbons}) of metrics.

\subsection{Gibbons solution in dS}

We assume the following ansatz ($\epsilon=-1$)
\begin{subequations}
\label{GibbonsdS}
\begin{align}
\D s^2 =& - e^{-H_{\rm G}/a(\tau)^{n-3}} \D \tau^2+a(\tau)^2 e^{H_{\rm G}/[(n-3)a(\tau)^{n-3}]}
h_{IJ}(y)\D y^I \D y^J\,, \\
\phi=& \sqrt{-\epsilon \frac{(n-2)}{4(n-3)\kappa_n}}\frac{H_{\rm G}}{a(\tau)^{n-3}}\,, 
\end{align}
\end{subequations}
where $a(\tau)=e^{H\tau}$,  $H_{\rm G}=H_{\rm G}(y^I)$ with $\Delta_h H_{\rm G}=0$, 
and $h_{IJ}$ is a $\tau$-independent Ricci flat metric. $H$ is the Hubble parameter and the $H=0$ case reduces to the 
Gibbons solution (\ref{Gibbons}).
Substituting the above ansatz into the field equations obtained 
from the action (\ref{action:pot}), the potential is 
\begin{align}
\label{Pot:Gibbons}
V(\phi)=\frac{H^2}{\kappa_n} e^{2\sqrt{(n-3)\kappa_n/(n-2)} \phi }\left[
\frac 12 (n-1)(n-2)-(n-1)\sqrt{(n-2)(n-3)\kappa_n}\phi+(n-3)(n-2)\kappa_n\phi^2
\right]\,,
\end{align}
which is expressed as (\ref{pot:superpotential}) in terms of the  superpotential as 
\begin{align}
\label{}
W(\phi)=\frac{i H}{2\sqrt{\kappa_n}} 
e^{\sqrt{(n-3)\kappa_n/(n-2)} \phi } \left(
\sqrt{\frac{(n-3)\kappa_n}{n-2}} \phi-1 
\right) \,. 
\end{align}
The origin $\phi=0$ is the unique critical point of the superpotential.
The another critical point of the potential $\phi=1/\sqrt{(n-2)(n-3)\kappa_n}$ does not 
extremize the superpotential and is irrelevant for the present analysis. 

Contrary to the Fisher solution in dS, this solution (\ref{GibbonsdS}) is generically dynamical, since the solution fails to admit any additional symmetry like (\ref{taurho}). In this dynamical form,  one cannot obtain the asymptotically AdS solution by the Wick rotation $H\to -i\ell^{-1}$. 
This is reminiscent of the Kastor-Traschen solution \cite{Kastor:1992nn} in the Einstein-Maxwell-$\Lambda(>0)$ system.
Kastor and Traschen have demonstrated that the Majumdar-Papapetrou multi-center solution 
\cite{Majumdar:1947eu,Papaetrou:1947ib} can be 
embedded into dS universe at the expense of introducing the positive cosmological constant \cite{Kastor:1992nn}. Their solution is 
specified by a harmonic function despite the explicit time dependence. 
If the harmonic function has the multiple point sources,  the Kastor-Traschen solution describes multiple colliding black holes in contracting dS universe ($H<0$) or multiple splitting white holes in expanding dS universe ($H>0$).
In contrast, the case with a single point source culminates in the lukewarm limit of the Reissner-Nordst\"om-dS solution, which admits a static patch \cite{Brill:1993tm}. 

Here we argue that an analogous situation happens also for the Gibbons solution in dS universe
(\ref{GibbonsdS}). As it turns out, static Killing vector emerges by restricting to the spherically symmetric case. In the appendix, we also present a dynamical generalization of the Gibbons solution into the asymptotically FLRW universe, in lieu of the dS universe.

\subsection{Ellis-Gibbons solution in dS: plus branch}

For an exceptional case, let us consider a monopole source
$H_{\rm G}=M/\rho^{n-3}$ in a flat space $h_{IJ}\D y^I \D y^J= \D \rho^2+\rho^2 \D \Omega_{n-2}^2$, for which  the solution is unchanged under $\tau\to \tau+c$, $\rho\to \rho e^{-Hc}$.
In this case, one can find the static patch, which is achieved by 
\begin{align}
\label{}
r=\rho e^{H \tau} \,, \qquad 
t= \tau+ \int \frac{H r e^{(n-2)M/[(n-3)r^{n-3}]}}{F^+_{\rm G}( r)}\D  r\,,
\end{align}
where 
\begin{align}
\label{FG1}
F_{\rm G}^+(r)\equiv  1 - H^2 r^2 e^{(n-2)M/[(n-3) r^{n-3}]}\,, 
\end{align}
yielding the static solution in the form
\begin{subequations}
\label{GibbonsdSstatic1}
\begin{align}
\D s^2&=-e^{-M/ r^{n-3}} F_{\rm G}^+( r)\D t^2+e^{M/[(n-3) r^{n-3}]}
\left(
\frac{\D  r^2}{F^+_{\rm G}( r)}+ r^2 \D \Omega^2_{n-2} 
\right)\,, \\
\label{GibbonsdSstaticscalar1}
\phi=& +\sqrt{-\epsilon \frac{(n-2)}{4(n-3)\kappa_n}}\frac{M}{ r^{n-3}}\,.
\end{align}
\end{subequations}
This is a natural generalization of the Ellis-Gibbons solution (\ref{EllisGibbons})
in dS universe. Note that the appealing property of linearization inherent to the 
Gibbons solution is now marred in this static form. In the following analysis, 
we assume $H>0$.

As established in our previous paper \cite{paperI}, the asymptotically flat Gibbons solution is 
necessarily singular at $r=0$, which corresponds to the scalar curvature singularity only for 
$M<0$. In contrast, the corresponding dS solution 
inescapably possesses the scalar curvature singularity at $r=0$, which can be deduced by 
the behavior of the Ricci scalar 
\begin{align}
\label{}
R= -\frac {(n-2)(n-3)M^2}{4r^{2(n-2)}} e^{-M/[(n-3)r^{n-3}]}+(n-1)H^2 e^{M/r^{n-3}}
\left(n-\frac{nM}{r^{n-3}}+\frac{3(n-2)M^2}{4r^{2(n-3)}}\right)\,. 
\end{align}
This blows up at $r=0$ insensitive to the sign of $M$ for $H\ne 0$. 

Let us look into the causal nature of this singularity. 
Since $F_{\rm G}^+(r)$ does not enter in the expression for the affine parameter of the radial null geodesics, the singularity $r=0$ locates at an infinite affine distance for $M<0$ in $n\ge 5$, and at finite affine distance otherwise. To see the signature of the $r=0$ surface, the following criterion is of use
\begin{align}
\label{criterionGibbons}
\lim _{r\to 0+}\int^r r^p e^{q/r^{n-3}} \D r \to \left\{
\begin{array}{cc}
 +\infty     &  (q>0)\,,  \\
  0    &   (q<0)\,,
\end{array}
\right.
\end{align}
where $p$ and $q$ are constants. 
From $r_*=\int (e^{(n-2)M/[2(n-3)r^{n-3}]}/F_{\rm G}^{+})\D r\to 0\mp$ 
as $r\to 0$ for $M\gtrless 0$, 
the singularity at $r=0$ is spacelike for $M>0$, while it is timelike for $M<0$.

Contrary to the $H=0$ case, 
the singularity at $r=0$ can be covered by an event horizon of a black hole, 
provided that the mass parameter $M$ satisfies 
\begin{align}
\label{GibbonsdSplus:horizon}
0<M < \frac{2H^{3-n}}{(n-2)e}\,.
\end{align}
Under this condition, 
the equation $F_{\rm G}^+(r)=0$ admits two real distinct roots $r_+$ and $r_{\rm c}(>r_+>0)$, 
which stand for respectively the event horizon of a black hole and the cosmological horizon. 
Consequently, the global structure of the solution under (\ref{GibbonsdSplus:horizon}) 
is the same as the Schwarzschild-dS black hole (see figure \ref{fig:PDFdS}).

Leaving aside the the regularity of these horizons, we have to pay some attention to the asymptotic structure of the spacetime. At fist glance, it may be conceivable that the metric (\ref{GibbonsdSstatic1}) approaches to the dS universe as $r\to \infty$.  In spite of this, the $r>r_{\rm c}$ region is highly dynamical, so that the notion of ``asymptotically dS'' is rather ambiguous, as opposed to the $\rho \to \infty$ limit of the McVittie's form (\ref{GibbonsdS}). 
Since the static solution (\ref{GibbonsdSstatic1}) does not look like the dS metric around the cosmological horizon,  it is not sure whether the solution (\ref{GibbonsdSstatic1}) falls into the 
well-defined framework of asymptotic dS family.

\subsection{Ellis-Gibbons solution in AdS: plus branch}

Since the Hubble parameter enters in the metric (\ref{FG1}) in the quadratic form, 
one can simply Wick rotate $H\to -i \ell^{-1}$ to give an asymptotically AdS static solution (\ref{GibbonsdSstatic1}). 
The function $F_{\rm G}^+(r) $ in (\ref{FG1}) is now replaced by by 
\begin{align}
\label{}
F_{\rm G}^+(r)= 1 + (r^2/\ell^2) e^{(n-2)M/[(n-3) r^{n-3}]}>0\,, 
\end{align}
implying the nonexistence of the horizon.  

The $r=0$ surface has a timelike structure, which becomes null infinity for $M<0$ with $n\ge 5$.  
The metric therefore describes a naked singularity at $r=0$ in asymptotically AdS spacetime.

\subsection{Ellis-Gibbons solution in dS: minus branch}

As in the Fisher solution in dS space (\ref{JNWdS:static2}), we try to look for a new solution, 
by leaving $F_{\rm G}^+( r)$ arbitrary, taking a different sign in the scalar field (\ref{GibbonsdSstaticscalar1}),
 while maintaining the metric form (\ref{GibbonsdSstatic1}). 
This prescription works again only in $n=4$  and reads
\begin{subequations}
\label{GibbonsdSstatic2}
\begin{align}
\D s^2&=-e^{-M/ r}  F^-_{\rm G}( r)\D t^2+e^{M/ r}
\left(
\frac{\D r^2}{F^-_{\rm G}( r)}+ r^2 \D \Omega_{ 2} ^2
\right)\,, \\
\label{GibbonsdSstaticscalar2}
\phi=&- \sqrt{ \frac{-\epsilon}{2\kappa_4}}\frac{M}{ r}\,.
\end{align}
\end{subequations}
where $F_{\rm G}^-( r)$ is fixed by Einstein's equations with the potential (\ref{Pot:Gibbons}) to be 
\begin{align}
\label{FGdSm}
F_{\rm G}^-( r)\equiv 1-H^2(2M^2+2M  r+ r^2) \,. 
\end{align}
We call (\ref{GibbonsdSstatic2}) as the minus-branch solution and set $H>0$ hereafter.

The coordinate $r$ corresponds to the affine parameter of the radial null geodesics. 
The curvature invariants for the above solution is singular at $r=0$ only for  $M<0$. 
But in either sign of $M$, the $r=0$ deserves a  p.p curvature singularity because of the divergence of 
$R_{\mu\nu}k^\mu k^\nu =-M^2/(2r^4)$. 
Inspecting (\ref{criterionGibbons}), the $r=0$  surface is null for $M>0$, spacelike for $M<-1/(\sqrt 2 H)<0$ and 
timelike for $-1/(\sqrt 2 H)<M<0$.

The minus branch solution (\ref{GibbonsdSstatic2}) also  admits a 
parameter range under which the event horizon exists for $r>0$, 
leading to
\begin{align}
\label{Gibbonsmassrange2}
-\frac{1}{H} < M < -\frac 1{\sqrt 2 H} \,. 
\end{align}
Namely, only the negative mass parameter provides the geometry of a black hole
with an event horizon at $r_+=-M-H^{-1}\sqrt{1-H^2M^2}>0$ and a cosmological horizon
at $r_c=-M+H^{-1}\sqrt{1-H^2M^2}>r_+$. Under the parameter range (\ref{Gibbonsmassrange2}),
the singularity at $r=0$ is spacelike, so that the Penrose diagram is identical to that of the Schwarzchild-dS black hole 
as displayed in figure \ref{fig:PDFdS}.

When the parameters take values out of the range (\ref{Gibbonsmassrange2}), 
the scalar/p.p curvature singularity at $r=0$ is visible. 
The spacetime with $|M|\ge H^{-1}$ is dynamical due to $F^-_{\rm G}(r)\le 0$, 
in the same spirit as the interior  of the Schwarzschild black hole. 
For $-(\sqrt 2 H)^{-1} <M<H^{-1}$ with $M\ne 0$, the spacetime admits 
only the cosmological horizon at  $r_c=-M+H^{-1}\sqrt{1-H^2M^2}$ in the 
$r>0$ region.

\subsection{Ellis-Gibbons solution in AdS: minus branch}

The asymptotic AdS solutions is merely realized by $H\to -i \ell$ in (\ref{GibbonsdSstatic2}), 
giving 
\begin{align}
\label{}
F_{\rm G}^-( r)= 1+\ell^{-2}(2M^2+2M  r+ r^2) >0\,,
\end{align}
in  (\ref{FGdSm}). 
This solution has therefore a null (timelike) singularity at $r=0$ 
for $M>0$ ($M<0$), which is not veiled by horizons.

\section{Ellis-Bronnikov class of solutions in (A)dS}
\label{sec:EG}

Lastly, we explore the Ellis-Bronnikov class. 
Contrary to the Fisher and Gibbons class, 
it turns out that the Ellis-Bronnikov solution allows the parameter range under which 
the solution describes the traversable wormhole.

\subsection{Ellis-Bronnikov solution in dS: plus branch}

We make a following ansatz ($\epsilon=-1$)
\begin{subequations}
\label{dSwormhole}
\begin{align}
\D s^2&=- e^{-2\beta U(\tau, \rho)}\D \tau^2+ a^2(\tau) e^{2\beta U(\tau, \rho)/(n-3)}\left(
1+\frac{M^2}{16(a(\tau)\rho)^{2(n-3)}} 
\right)^{2/(n-3)}(\D \rho^2+\rho^2 \D \Omega_{n-2}^2)\,, \\
\phi&=+\sqrt{-\epsilon\frac{(n-2)(1+\beta^2)}{\kappa_n(n-3)}} U(\tau,\rho)\,, 
\end{align}
\end{subequations}
where  $a(\tau)\equiv  e^{H \tau} $ and 
\begin{align}
\label{}
U(\tau,\rho)\equiv  \arctan \left(\frac{M}{2(a(\tau)\rho)^{n-3} [1-M^2/(16(a(\tau)\rho)^{2(n-3)})]}\right)\,. 
\end{align}
The $H=0$ recovers the Ellis-Bronnikov solution  in the isotropic coordinates (\ref{Ellis-Bronnikov:iso}).
For $\rho \to \infty$, the metric approaches to the dS universe with Hubble parameter $H$. 
Substituting the above ansatz to the field equations derived from the action (\ref{action:pot}), 
the potential of the scalar field is found to be
\begin{align}
\label{EG:pot}
V(\phi)=\frac{n-2}{2\kappa_n}H^2 e^{2\beta  \beta_0 \phi}
\Bigl[(n-2)(1+\beta^2)+(1-(n-2)\beta^2)\cos (2\beta_0 \phi)-(n-1)\beta
\sin (2\beta_0\phi)\Bigr]\,, 
\end{align}
where 
\begin{align}
\label{}
 \beta_0 \equiv  \sqrt{\frac{(n-3)\kappa_n }{(n-2)(1+\beta^2)}}\,. 
\end{align}
Also in this case, one can express the scalar potential (\ref{EG:pot}) as (\ref{pot:superpotential}) 
in terms of the following  superpotential 
\begin{align}
\label{}
W(\phi)=\frac{iH}{2\sqrt{\kappa_n}} e^{\beta \beta_0 \phi}\Big(\cos (\beta_0 \phi)-\beta \sin(\beta_0 \phi)\Big) \,.
\end{align}
This superpotential has a striking resemblance to (\ref{JNW:superpotential}). 
Note that $\beta $ is now entitled as a constant parameterizing the theory, instead of the 
solution parameter. One sees that the potential admits an infinite number of critical points
$\phi_{(k)}$ and $ \hat \phi_{(k)}$  ($k\in \mathbb Z$), where 
\begin{align}
\label{phik}
\phi_{(k)}\equiv \sqrt{\frac{(n-2)(1+\beta^2)}{(n-3)\kappa_n}} k \pi \,, \qquad 
\hat\phi_{(k)}\equiv \sqrt{\frac{(n-2)(1+\beta^2)}{(n-3)\kappa_n}}\left[{\rm arctan}\left(\frac{1}{(n-2)\beta}\right)+k \pi\right]\,.
\end{align}
 We can confirm that the former extrema $\phi=\phi_{(k)}$ of the potential also correspond to the 
critical points of the superpotential. This includes the origin $\phi=0$ of the scalar field, to which our solution approaches asymptotically. A simple calculation reveals that the value of the potential at these critical points are related by
\begin{align}
\label{}
\frac{V(\phi_{(k)})}{V(0)}= e^{2\pi k \beta} \,, \qquad 
\frac{V(\hat \phi_{(k)})}{V(0)}=\frac{n-3}{n-1}e^{2\pi k \beta +2\beta \arctan[1/((n-2)\beta)]}\,.
\end{align}

The solution (\ref{dSwormhole}) again admits a Killing vector $\xi^\mu=(\partial/\partial \tau)^\mu-H \rho (\partial/\partial \rho)^\mu$, 
since it is invariant under $\tau\to \tau+c$, $\rho\to \rho e^{-Hc}$. 
Defining 
\begin{align}
\label{}
\hat \rho=\rho e^{H\tau} \,, \qquad 
t=\tau+\int \frac{ H \hat \rho \hat V(\hat \rho)^{2/(n-3)}
e^{2\beta (n-2)\hat U(\hat \rho)/(n-3)}
}{f_{\rm EB}(\hat \rho)}\D \hat \rho\,, 
\end{align}
with 
\begin{subequations}
\begin{align}
\label{}
f_{\rm EB}(\hat \rho)\equiv &\, 1 - H^2 \hat \rho^2 \hat V(\hat \rho)^{2/(n-3)} e^{2(n-2)\beta \hat U(\hat \rho)/(n-3)} 
\,, \\
\hat U(\hat \rho)\equiv &\, \arctan \left(\frac{M}{2\hat \rho^{n-3}[1-M^2 /(16 \hat \rho^{2(n-3)})]}\right)\,, \\
\hat V(\hat \rho)\equiv &\,  1+\frac{M^2}{16 \hat \rho^{2(n-3)}}\,, 
\end{align}
\end{subequations}
one arrives at a static form
\begin{subequations}
\begin{align}
\label{}
\D s^2 &= -e^{-2\beta \hat U(\hat \rho)} f_{\rm EB}(\hat \rho)\D  t^2 +\hat V(\hat \rho)^{2/(n-3)}
e^{2\beta\hat U(\hat \rho)/(n-3)}
\left(\frac{\D \hat \rho^2}{f_{\rm EB}(\hat \rho)}
 + \hat \rho^2 \D \Omega_{n-2}^2 \right)\,, \\
\phi&=\sqrt{-\epsilon\frac{(n-2)(1+\beta^2)}{\kappa_n(n-3)}} \hat U(\hat \rho) \,. 
\end{align}
\end{subequations}
By the further coordinate transformation
\begin{align}
\label{}
\hat \rho^{n-3}=\frac 12 r^{n-3} \left(1+\sqrt{1+\frac{M^2}{4r^{2(n-3)}}}\right) \,, 
\end{align}
one can cast the above solution into a simpler form
\begin{subequations}
\label{EBdSp}
\begin{align}
\D s^2=&-e^{-2\beta U(r)}F_{\rm EB}^+(r)\D t^2% \notag \\ &
+V(r)^{1/(n-3)}e^{2\beta U(r)/(n-3)}
\left(\frac{\D r^2}{V(r)F_{\rm EB}^+(r)}+r^2 \D \Omega_{n-2}^2\right)\,, 
\\
\phi=&+\sqrt{-\epsilon\frac{(n-2)(1+\beta^2)}{\kappa_n(n-3)}}U(r) \,, 
\end{align}
\end{subequations}
where 
$U(r)=\arctan({M}/{2r^{n-3}})$, 
$V(r)=1+{M^2}/(4r^{2(n-3)})$ as before (\ref{UV}), and 
\begin{align}
\label{EBdSFp}
F_{\rm EB}^+(r)&\equiv 1-H^2 r^2 V(r)^{1/(n-3)}
e^{2(n-2)\beta U(r)/(n-3)}\,. 
\end{align}
Here use has been made of 
\begin{align}
\label{}
\hat U(\hat\rho)=U(r) \,, \qquad 
\hat \rho^2 \hat V(\hat \rho)^{2/(n-3)} =r^2 V(r)^{1/(n-3)} \,, \qquad 
\hat V(\hat \rho)^{2/(n-3)}\D \hat \rho^2=V(r)^{-(n-4)/(n-3)}\D r^2 \,. 
\end{align}
The  metric (\ref{EBdSp}) can be viewed as the Ellis-Bronnikov solution in the dS universe, 
since the $M=0$ case reduces to the dS spacetime, while $H=0$ recovers  the Ellis-Bronnikov solution. The solution (\ref{EBdSp})  is invariant under the simultaneous sign change (\ref{Mbeta}) of 
$(M, \beta)$, which permits one to get centered on the $M>0$ case. 
From now on, we set $H>0$. 

%The Ellis-Bronnikov solution (\ref{Ellis-Bronnikov}) describes a regular wormhole solution 
%for any parameter regions. For $H>0$, the solution  (\ref{EBdSp})  may admit an event horizon of the black hole. 

Let us first examine the coordinate singularity $r=0$. 
We observe that the $r=0$ surface is completely regular. For instance the Ricci scalar 
is kept finite at $r=0$,
\begin{align}
\label{}
R=&-\frac{(n-2)(n-3)}{4r^{2(n-2)}V(r)^{(n-2)/(n-3)}}(1+\beta^2) e^{-2\beta U(r)/(n-3)}
\notag \\
&+\frac{(n-1)H^2}{r^{2(n-3)}V(r)}e^{2\beta U(r)}\left[n r^{2(n-3)}-nM\beta r^{n-3}
+\frac 14 M^2 \left(2(n-3)+3(n-2)\beta^2\right)\right]\,. 
\end{align}
One can verify by rudimentary calculations that any Riemann tensor components in a frame parallelly propagated along the radial null geodesics are nondiverging at $r=0$. It follows that $r=0$ is merely a coordinate singularity.  
The extension across  $r=0$ can be achieved by the replacement 
of $U(r)$ by $U(r)\to \pi/2-\arctan(2r^{n-3}/M)$ with $x=r^{n-3}$. 
In this coordinate system, the solution reads 
\begin{subequations}
\label{EBdSpx}
\begin{align}
\D s^2=&-e^{-2\beta U_x(x)}F_x^+(x)\D t^2
+V_x(x)^{1/(n-3)}e^{2\beta U_x(x)/(n-3)}
\left(\frac{\D x^2}{(n-3)^2 V_x(x)F_{x}^+(x)}+ \D \Omega_{n-2}^2\right)\,, 
\\
\phi=&+\sqrt{-\epsilon\frac{(n-2)(1+\beta^2)}{\kappa_n(n-3)}}U_x(x) \,, 
\end{align}
\end{subequations}
where 
\begin{subequations}
\begin{align}
\label{}
U_x(x)\equiv \,&\frac{\pi}{2}-\arctan\left(\frac{2x}{M}\right)\,, \\
V_x(x)\equiv \,& x^2+\frac{M^2}{4} \,, \\
F_x^+(x)\equiv\, &1-H^2 V_x(x)^{1/(n-3)}
e^{2(n-2)\beta U_x(x)/(n-3)}\,. 
\end{align}
\end{subequations}
One finds that every component of the metric and its inverse
is now smooth for any $x\in (-\infty, \infty)$. It follows that the metric (\ref{EBdSpx}) yields the maximal extension of the spacetime. 
The solution is geodesically complete and admits no scalar/p.p curvature singularities. 

\begin{figure}[t]
\begin{center}
\includegraphics[width=7cm]{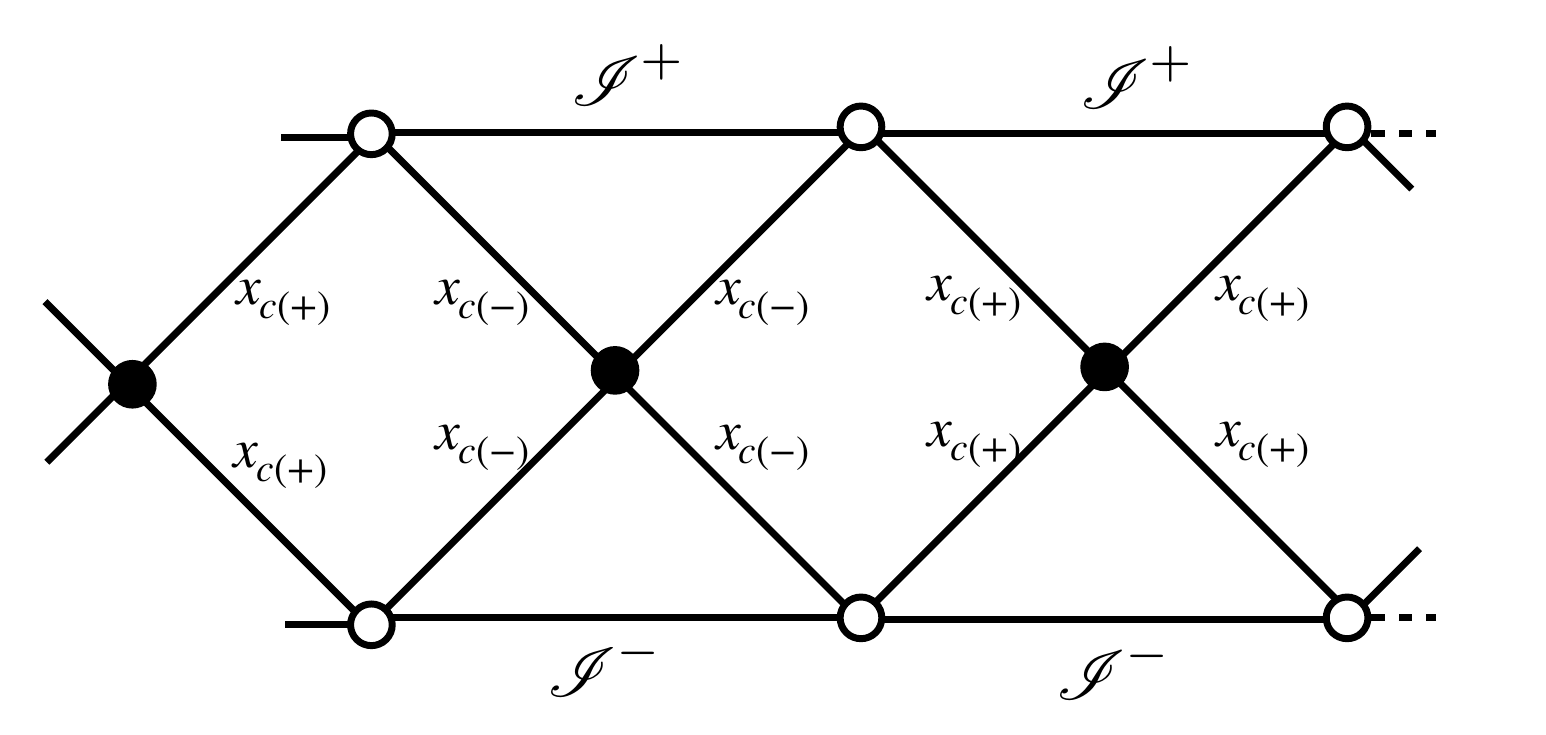}
\caption{A possible conformal diagram for the dS Ellis-Bronnikov wormholes
(plus (\ref{EBdSpx}) and minus (\ref{EBdS2}) branches). The required parameter range is 
(\ref{EBdSplus:horizon}) and (\ref{EBdSminus:horizon}), respectively. 
}
\label{fig:dSWH}
\end{center}
\end{figure}

We now argue that there exists a parameter range under which  the solution describes a 
dS wormhole. 
Let us first note that the solution allows two distinct roots 
$x_{c(\pm)}$ for $F_{\rm EB}^+(x_{c(\pm)})=0$, provided $M(>0)$ satisfies the 
following inequality
\begin{align}
\label{EBdSplus:horizon}
0< M <\frac{2H^{3-n}}{\sqrt{1+(n-2)^2\beta^2}}e^{-(n-2)\beta (\pi/2-\arctan[(n-2)\beta])}\,. 
\end{align}
Both of these roots $x_c^{(\pm)}$ should be regarded 
as the cosmological horizons. 
Hence, the solution satisfying (\ref{EBdSplus:horizon}) is identified as a 
static wormhole in dS universe. 
The Penrose diagram is shown in figure \ref{fig:dSWH}.

If $M$ is too large than the critical value in (\ref{EBdSplus:horizon}), $F^+_{x}(x)$ 
is negative definite throughout $x\in \mathbb R$. This is dynamical in the same spirit as the interior of the Schwarzschild black hole. 

A local characterization of the wormhole is the existence of the throat \cite{Morris:1988cz}, corresponding to 
the critical point of the areal radius $S\equiv V_x(x)^{1/[2(n-3)]}e^{\beta U_x(x)/(n-3)}$. 
In the present case, this locates at 
\begin{align}
\label{throat}
x=x_{\rm th}=\frac 12 M \beta \,.
\end{align}
This does not coincide with the coordinate boundary $r=0$ for the original metric (\ref{EBdSp}) 
when $\beta \ne 0$. 
As emphasized in our previous paper \cite{paperI},  
the analysis of geodesics is more important than this local definition, 
since the wormhole is a global concept. 
Our argument above explicitly shows that there exist null geodesics emanating from 
past infinity ($t=-\infty$ and $r\to \infty$) which can arrive at distinct future null infinity
($t=+\infty$ and $x\to \infty$) or ($t=+\infty$ and $x\to- \infty$), 
illustrating that the  solution (\ref{EBdSpx}) indeed describes a wormhole.

\subsection{Ellis-Bronnikov solution in AdS: plus branch}

Upon the Wick rotation $H\to -i \ell^{-1}$, the solution (\ref{EBdSp}) becomes asymptotically AdS
as $r\to 0$. Since the nonsingular nature  of $r=0$ does not change compared to the dS case, 
the solution can be recast into the form (\ref{EBdSpx}) with $x\in \mathbb R$, where $F_x^+(x)$ 
 is now
\begin{align}
\label{EBAdS1}
F_x^+(x)= 1+\frac{1}{\ell^2} V_x(x)^{1/(n-3)}
e^{2(n-2)\beta U_{x}(x)/(n-3)} >0 \,. 
\end{align}
Since there appear no scalar/p.p curvature singularities in the entire parameter region, 
the solution indeed describes a wormhole in AdS.   The throat exists at (\ref{throat}).

In the $x\to \infty$ region, the scalar field behaves as $\phi \to 0$. 
The potential $V(\phi)$ at the origin has the mass spectrum 
$m^2 =\epsilon V''|_{\phi=0, \epsilon=-1}=-2(n-3)/\ell^2$,\footnote{The $\epsilon $ term in front of $V''$ is deduced from the multi-dimensional covariant form of the mass matrix
$(m^2)^i{}_j =G^{ik}\partial_k\partial_j V$ at the critical point $\partial_i V=0$, where $G_{ij}$ is the scalar metric of the scalar fields normalized by $\ma L_{\rm pot}=\frac 12 G_{ij}(\phi^k) \nabla_\mu \phi^i \nabla^\mu \phi^j -V(\phi^k)$. The same remark applies to the $\epsilon$ term in the potential 
(\ref{pot:superpotential}). 
} which never underruns the 
Breitenlohner-Freedman bound $m_{\rm BF}^2=-(n-1)^2/(4\ell^2)$ \cite{Breitenlohner:1982jf}
since $m^2-m_{\rm BF}^2=(n-5)^2/(4\ell^2)\ge 0$. 
Note that this value of mass lies in the unitary range $m_{\rm BF}^2\le m^2 \le m_{\rm BF}^2+\ell^{-2}$
for $4\le n\le 7$,  for which the scalar field exhibits a slow fall-off at infinity. 
In this case, the standard definition of globally conserved mass 
\cite{Abbott:1981ff,Ashtekar:1984zz,Henneaux:1985tv,Katz:1996nr,Hollands:2005wt}
should be modified. Following the prescription in
 \cite{Hertog:2004dr,Henneaux:2006hk}, 
the mass in the $x>0$ region reads
\begin{align}
\label{}
\ma M_{x>0}=\frac{(n-2)\Omega_{n-2}}{2\kappa_n} M \beta \,.
\end{align}
In an analogous fashion, the mass $\ma M_{x<0}$ and the AdS radius $\ell_{x<0}$
in the $x<0$ region are given by
\begin{align}
\label{}
\ma M_{x<0}=-\frac{(n-2)\Omega_{n-2}}{2\kappa_n} e^{\beta \pi}M \beta \,, \qquad 
\ell_{x<0}=-e^{-\pi \beta } \ell \,. 
\end{align}
It follows that the $\beta \ne 0$ case joints the universes of different masses
with the opposite sign.

In the $x\to - \infty $ limit, the scalar field converges to the asymptotic value 
\begin{align}
\label{EBplus:vac}
\phi(x)\to \phi_{(1)}\quad (x\to -\infty)\,,
%\phi_{-\infty} \equiv \pi \sqrt{\frac{(n-2)(1+\beta^2)}{(n-3)\kappa_n}} \,. 
\end{align}
where $\phi_{(1)}$ is given by (\ref{phik}). 
At this critical point, the mass eigenvalue normalized by the AdS length is the same as the one at the origin
$m^2|_{\phi=\phi_{(1)}}=-2(n-3)/\ell_{x<0}^2$. 
Observe that this is also the critical point of the potential and the superpotential
 $V'(\phi_{(1)})=W'(\phi_{(1)})=0$. This underlines the solitonic 
 property of our wormhole solution, connecting two disjoint vacua. 

\begin{figure}[t]
\begin{center}
\includegraphics[width=4cm]{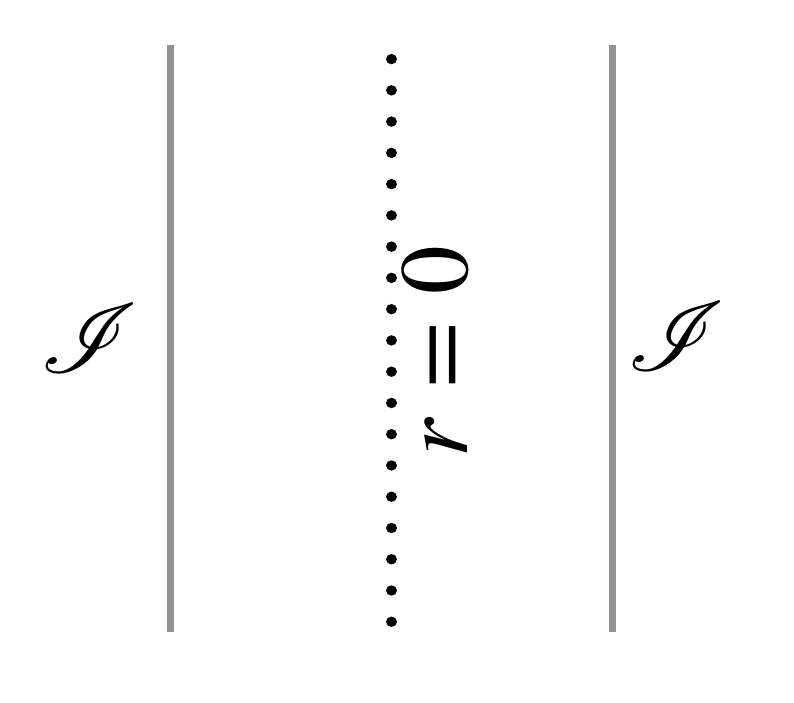}
\caption{A possible conformal diagram for the AdS Ellis-Bronnikov wormholes
(plus (\ref{EBAdS1}) and minus (\ref{FEBAdS2}) branches). 
}
\label{fig:AdSWH}
\end{center}
\end{figure}

The global structure of the AdS Ellis-Bronnikov solution is shown in figure~\ref{fig:AdSWH}.

\subsection{Ellis-Bronnikov solution in dS: minus branch}

Lastly, we try to seek a new flipped solution of the Ellis-Bronnikov solution. 
As it turns out, this prescription only works in $n=4$, yielding 
\begin{subequations}
\label{EBdS2}
\begin{align}
\D s^2=&-e^{-2\beta U(r)}F^-_{\rm EB}(r)\D t^2
% \notag \\ &
+V(r)e^{2\beta U(r)}
\left(\frac{\D r^2}{V(r)F^-_{\rm EB}(r)}+r^2 \D \Omega_{n-2}^2\right)\,, 
\\
\phi=&-\sqrt{-\epsilon\frac{2(1+\beta^2)}{\kappa_4}}U(r) \,, 
\end{align}
\end{subequations}
where $U(r)$ and $V(r)$ are given by (\ref{UV})  respectively. 
Inserting above into field equations with a potential given by 
(\ref{EG:pot}), one can determine $F_{\rm EB}^-(r)$ to be 
\begin{align}
\label{FEBdS}
F_{\rm EB}^-(r)&=1-H^2 \left(
r^2+2M\beta r+M^2\frac{1+8\beta^2}{4}
\right)\,.
\end{align}
One can deduce that the metric (\ref{EBdS2}) is invariant under $\beta\to -\beta$ and $M\to -M$ 
as in the $H=0$ case (\ref{Mbeta}), 
which enables one to concentrate on the $M>0$ case. 
Since the piece multiplied by a Hubble parameter in (\ref{FEBdS}) has a negligible contribution around $r=0$, 
$r=0$ is neither scalar nor p.p curvature singularity. Since $r $ plays the role of the affine parameter along the radial null geodesics, the $r=0$ surface is not null infinity. 
One can then extend the spacetime across $r=0$ by 
$U(r)\to  \pi/2-\arctan(2r/M)$ into the $r<0$ region.

In either sign of $\beta$, 
there appear the surfaces $F_{\rm EB}^-(r_c^{(\pm)})=0$ in $r\in \mathbb R$, 
provided 
\begin{align}
\label{EBdSminus:horizon}
0<M < \frac{2}{H\sqrt{1+4\beta^2}} \,.
\end{align}
These loci $r_c^{(\pm)}$ correspond to the two distinct cosmological horizons, 
rather than a black hole horizon. Under the parameter range (\ref{EBdSminus:horizon}), 
the solution then describes a wormhole connecting two dS universes with a 
throat at $r=r_{\rm th}=M\beta /2$. The Penrose diagram is identical to figure \ref{fig:dSWH}. 
When the parameter $M$ fails to satisfy the above criterion, 
we have a dynamical spacetime in the sense of  $F_{\rm EB}^-(r)<0$.

\subsection{Ellis-Bronnikov solution in AdS: minus branch}

By the Wick rotation $H\to -i \ell^{-1}$ in (\ref{EBdS2}), one gets the asymptotically AdS solution 
where $F_{\rm EB}^-(r)$ is now understood to be 
\begin{align}
\label{FEBAdS2}
F_{\rm EB}^-(r)&=1+\frac{(1+4\beta^2)M^2}{4\ell^2}+\frac{(r+M\beta)^2}{\ell^2} \,, 
\end{align}
which is obviously positive. One verifies that the $r=0$ surface is not the scalar/p.p curvature singularity nor null infinity, and that it has a timelike structure. It follows that one can extend the spacetime into the $r<0$ region by $U(r) \to \pi/2 -\arctan(2r/M)$. 
The resulting spacetime is geodesically complete and describes a regular wormhole in AdS
 with a throat at $r=r_{\rm th}=M\beta /2$. 

As in the plus branch solution, one can compute the 
mass in the $r>0$ region as 
\begin{align}
\label{}
\ma M_{r>0}=\frac{4\pi}{\kappa_4}M\beta\left(1+\frac{M^2}{3\ell^2} (1+4\beta^2)\right)\,.
\end{align}
The mass and the AdS radius in the $r<0$ region are 
\begin{align}
\label{}
\ma M_{r<0}=-\frac{4\pi e^{\pi \beta}}{\kappa_4}M\beta\left(1+\frac{M^2}{3\ell^2} (1+4\beta^2)\right)\,, 
\qquad \ell_{r<0}=- \ell e^{\pi \beta} \,. 
\end{align}
Thus the $\beta \ne 0$ case yields the asymmetric extension. 
The asymptotic value of the scalar field in the $r\to \pm \infty$ is 
\begin{align}
\label{}
\phi \to 0 \quad (r\to \infty)\,, \qquad 
\phi \to \phi_{(-1)} \quad (r\to -\infty)\,,
\end{align}
where $\phi_{(-1)}$ is given by (\ref{phik}). 
This critical point also extremizes the superpotential. We note that this extremum 
$\phi_{(-1)}$ is different from the one $\phi_{(+1)}$ for the plus-branch (\ref{EBplus:vac}).
It turns out that the minus-branch solution also interpolates two different AdS vacua.

\section{Summary and concluding remarks}
\label{sec:conclusion}

This paper is intended to shed some new insight into the problem for the construction of exact solutions in (A)dS.
We have presented a useful method to generate static solutions in the Einstein-phantom-scalar system with a potential. Interestingly, all the potentials are expressed in terms of the superpotential as (\ref{pot:superpotential}). This is fairly nontrivial since the existence of this kind of superpotential is not guaranteed in general. 

Based on our proposed methodology, we have constructed the asymptotically (A)dS versions of (i) the Fisher solution, (ii) the Gibbons solution, and (iii) the Ellis-Bronnikov solution. 
For each class of solutions, there appear two distinct family of solutions (plus and minus branches). 
Our main results are summarized in table \ref{fig:summary}. 
Contrary to the asymptotically flat case, the (A)dS Fisher and Gibbons solutions may admit a black hole horizon which covers the central singularity. We have clarified the precise conditions under which each solution possesses the event/cosmological horizons. One of the central results of the current paper is that both branches of the dS Ellis-Bronnikov solutions admit the parameter range under which the solutions correspond to traversable wormholes in dS universe, whereas both branches of the 
AdS Ellis-Bronnikov solutions always describe the regular wormholes in the entire parameter region. 
Interestingly, the extended spacetime admits another AdS vacuum at infinity, which also corresponds to the critical point of the superpotential. 
This provides an invaluable instance of wormholes in AdS, which would be fruitful for the holographic entanglement.

\begin{table}[t]
\begin{center}
\begin{tabular}{c||ccc}
class & branch & asymptotics  & configurations \\ 
\hline \hline 
\multirow{4}{*}{Fisher} 	
& \multirow{2}{*}{plus} & dS & black hole for (\ref{JNWdS:horizoncond1}) \\
& & AdS & naked singularity \\ \cline{2-4}
& \multirow{2}{*}{minus} & dS & black hole for (\ref{FdSminus:signature}) \\
& & AdS & black hole for (\ref{FAdSminus:horizon}) \\ \hline
\multirow{4}{*}{Ellis-Gibbons} 	
& \multirow{2}{*}{plus} & dS & black hole for (\ref{GibbonsdSplus:horizon})  \\
& & AdS & naked singularity  \\ \cline{2-4}
& \multirow{2}{*}{minus} & dS & black hole for (\ref{Gibbonsmassrange2}) \\
& & AdS & naked singularity \\ \hline
\multirow{4}{*}{Ellis-Bronnikov} 	
& \multirow{2}{*}{plus} & dS & wormhole for (\ref{EBdSplus:horizon}) \\
& & AdS &  wormhole \\ \cline{2-4}
& \multirow{2}{*}{minus} & dS &  wormhole for (\ref{EBdSminus:horizon}) \\
& & AdS &  wormhole \\ \hline
\end{tabular}
\caption{Asymptotically (A)dS solutions.}
\label{fig:summary}
\end{center}
\end{table}

For the generalization of each solution into (A)dS case, 
we have encountered two distinct family (plus and minus branches) of solutions. 
Since the scalar field potential is fixed, 
one may foresee that there exists a more general solution which unifies both of these branches. 
This can be validated only by a direct integration of Einstein's equations. 
Unfortunately this is a demanding job, since the nonlinearity still survives 
even for the spacetimes with spherical symmetry.  
We hope to return to this issue in the future.

The results of the present paper seem to be extended into many directions. As we argued in section~\ref{sec:Gibbons}, the spherical Gibbons solutions in dS space are eventually written into a static form. In contrast, the solution with multiple point sources (\ref{GibbonsdS}) is not static. A natural interpretation of this solution is the collision of point sources in the contracting universe ($H<0$), or splitting point sources in the expanding universe ($H>0$). An analogous dynamical solution in FLRW universe is presented in appendix. It is interesting to explore the explicit conditions under which this viewpoint is correct for these dynamical solutions.

Another direction of future work is to include the (dilatonic) charge. In our previous paper \cite{paperII}, the dilatonic charged static solutions satisfying asymptotic flatness have been constructed. By taking these solutions as a seed, one may be able to generate asymptotically (A)dS charged solutions. An important observation in the asymptotically flat case is that there exists a critical value of the coupling constant of the dilaton, at which the metric takes a considerably different form. It would be interesting to see how the coupling affects the physical/causal properties of asymptotically (A)dS solutions.

In our series of papers \cite{paperI,paperII}, we have not discussed the stability of the wormhole solutions.
Obviously, this is an important issue to be studied. The linear instability for asymptotically flat cases has been reported in \cite{Novikov:2009vn,Nandi:2016ccg,Cremona:2018wkj,Torii:2013xba}. It is worth investigating whether the electric charge or the scalar potential stabilizes the wormholes.

\subsection*{Acknowledgements}

This work  is partially supported by 
Grant-in-Aid for Scientific Research (A) from JSPS 17H01091 and (C) 20K03929.

\appendix
\renewcommand{\theequation}{A.\arabic{equation}}
\setcounter{equation}{0}

\section{Curvature decomposition}
\label{app:curv}

The time-dependent metric that we encountered in the main text takes the following universal form
\begin{align}
\label{}
\D s^2=-f(t,y) \D t^2+ a(t)^2 f(t,y)^{-1/(n-3)} h_{IJ}(y) \D y^I \D y^J \,, 
\end{align}
where the base space metric $h_{IJ}$ is assumed to be independent of the time coordinate $t$. 
This class of metric encompasses the spherical
 Fisher (\ref{JNWdyn}), Ellis-Gibbons (\ref{GibbonsdS}) and Ellis-Bronnikov (\ref{dSwormhole}) solutions  in dS universe, 
 as well as multi Gibbons  solution in dS universe (\ref{GibbonsdS}). 
 Besides, the asymptotic FRLW metric which will be discussed in appendix \ref{Gibbons:FLRW} also falls into the above class of solutions.

The connections are given by
\begin{align}
\label{}
\Gamma^t{}_{tt}&=\frac{\dot f}{2f} \,, \qquad \Gamma^t{}_{tI}=\frac{D_I f}{2f} \,, \qquad
 \Gamma^t{}_{IJ}=\frac F{f^{(n-2)/(n-3)}}h_{IJ} \,, \qquad 
\Gamma^I{}_{tt}=\frac {f^{1/(n-3)}}{2a^2}D^I f \,,\notag \\
\Gamma^I{}_{tJ}&=\frac{F}{a^2}\delta^I{}_J \,,\qquad 
\Gamma^I{}_{JK}=\, {}^{(h)}\Gamma^I{}_{JK}-\frac 1{2(n-3)f}\left(2\delta^I_{(J}D_{K)} f-h_{JK}D^I f\right)\,,
\end{align}
where the dot is the partial derivative with respect to $t$, 
$D_I$ is the covariant derivative of $h_{IJ}$ and we have denoted 
\begin{align}
\label{}
F \equiv a \dot a -\frac{a^2\dot f}{2(n-3)f} \,. 
\end{align}
The Riemann tensors are 
\begin{subequations}
\begin{align}
\label{}
R^{tI}{}_{tJ}=& \frac{f^{1/(n-3)}}{a^2}\left(\frac{(n-5)D^IfD_Jf}{4(n-3)f^2}-\frac{D^ID_Jf}{2f}+\frac{(Df)^2}{4(n-3)f^2}\delta^I{}_J\right)\notag \\
&+\frac 1{a^2 f}\left(\dot F-\left(\frac{(n-2)\dot f}{2(n-3)f}+\frac{\dot a}{a}\right)F\right)\delta^I{}_J\,, 
\\
R^{tI}{}_{JK}=&\frac{1}{a^2f^2}\left(2 fD_{[J}F-FD_{[J}f\right)\delta^I{}_{K]}\,, 
\\
R^{IJ}{}_{KL}=&\frac{f^{1/(n-3)}}{a^2}\Biggl({}^{(h)}R^{IJ}{}_{KL}
-\frac{(Df)^2}{2(n-3)^2 f^2}\delta^I{}_{[K}\delta^J{}_{L]} +\frac 2{(n-3)f}\delta^{[I}{}_{[K}D_{L]}D^{J]}f
\notag \\
&-\frac{2n-7}{(n-3)^2 f^2}\delta^{[I}{}_{[K}D_{L]}f D^{J]}f \Biggr)
+\frac{2F^2}{fa^4}\delta^I{}_{[K}\delta^J{}_{L]} 
\,.
\end{align}
\end{subequations}
The Ricci tensors are given by
\begin{subequations}
\begin{align}
\label{}
R^t{}_t=& \frac 1{2a^2 f^{(2n-7)/(n-3)}}\left((Df)^2-fD^2 f\right)+\frac{n-1}{a^2 f}\left[
\dot F-\left(\frac{(n-2)\dot f}{2(n-3)f}+\frac{\dot a}{a}\right)F
\right] \,, \\
R^t{}_I=&-\frac{n-2}{2(n-3)}\left[\frac 1f D_I(\partial_ t (\log f))
+\frac{(n-3)F}{a^2f^2}D_I f
\right]\,, \\
R^I{}_J=&\frac{f^{1/(n-3)}}{a^2}\left[{}^{(h)}R^I{}_J-\frac{1}{4(n-3)}\left(
(n-2)\frac{D^IfD_Jf}{f^2}-\frac{2D^2f}{f}\delta^I{}_J +\frac{2(Df)^2}{f^2}\delta^I{}_J
\right)\right]\notag \\
&+\frac 1{a^2 f}\left[
\dot F-\left(\frac{(n-2)\dot f}{2(n-3)f}+\frac{\dot a}{a}\right)F
+\frac{n-2}{a^2}F^2\right]\delta^I{}_J\,.
\end{align}
\end{subequations}

\section{Gibbons solution in expanding universe}

\label{Gibbons:FLRW}

In the main text, we have discussed dS generalization of the Fisher solution, the Gibbons solution and the
Ellis-Bronnikov solution. Suppose that the scale factor in the ansatz of the dS Fisher solution (\ref{JNWdyn}) is replaced by the power-law of the cosmic time. Although this metric is clearly asymptotically FLRW,  this spacetime fails to be a solution of Einstein-(phantom-)scalar system with a potential. This is indeed the case even for the case with a vanishing scalar field \cite{McVittie:1933zz}. Nevertheless, this prescription works for a certain functional form of scale factor as far as the Gibbons solutions are concerned. 

We make an ansatz for the solution ($\epsilon=-1$)
\begin{subequations}
\label{Gibbonstime}
\begin{align}
\D s^2 =&- e^{-(2ht +H_{\rm G})} \D t^2+e^{(2ht +H_{\rm G})/(n-3)} h_{IJ}\D y^I \D y^J \,, \\
\phi=&\sqrt{-{\epsilon}\frac{n-2}{4(n-3)\kappa_n}}(2ht +H_{\rm G}) \,,\qquad 
\Delta_{h} H_{\rm G}=0 \,, 
\end{align}
\end{subequations}
where $H_{\rm G}$ and $h_{IJ}$ are $t$-independent  and $h_{IJ}$ is an arbitrary Ricci flat metric. 
The parameter $h$ controls the expansion of the universe and 
$h=0$ reduces to the Gibbons solution (\ref{Gibbons}). 
Substituting the above ansatz into the field equations derived from the action (\ref{action:pot})
and using the curvature formula given in the previous appendix, 
one can determine the potential as the runaway form 
\begin{align}
\label{}
V(\phi)=\frac{(n-2)^2}{(n-3)^2\kappa_n}h^2 e^{2\sqrt{\kappa_n(n-3)/(n-2)}\phi}\,.
\end{align}
The  superpotential of the form (\ref{pot:superpotential}) with $\epsilon=-1$
is found to be 
\begin{align}
\label{}
W(\phi)=\frac{ih}{2(n-3)\sqrt{\kappa_n}} e^{\sqrt{\kappa_n(n-3)/(n-2)}\phi}\,. 
\end{align}
The potential above does not allow any critical points. 

Assuming $h>0$, we can define the cosmic time $\bar t$ by $e^{-h t}=-h \bar t$ ($-\infty <\bar t<0$). The metric is then brought into
\begin{align}
\label{}
\D s^2 =-e^{-H_{\rm G}} \D \bar t^2  +e^{H_{\rm G}/(n-3)}\bar a^2(\bar t)h_{IJ}\D y^I \D y^J\,,
\end{align}
where the scale factor is given by $\bar a(\bar t)=(-h \bar t)^{-1/(n-3)}$. 
Thus, the metric asymptotes to the expanding FLRW universe  as $r\to \infty$, provided $h_{IJ}=\delta _{IJ}$ and 
$H_{\rm G}=\sum_i M_i/|{\bf x}-{\bf x}_i|^{n-3}$. 
This solution does not admit additional Killing vectors even in the case with a single point source, so that it is genuinely dynamical. 
Similar dynamical solutions in FLRW universe can be found e.g., 
in \cite{Maeda:2009ds,Maeda:2010ja,Nozawa:2010zg,Chimento:2012mg,Klemm:2015qpi,Ishihara:2015hmx}, 
which describes respectively a single equilibrium black hole in expanding universe.
However, the present solution differs from these solutions in that the current metric (\ref{Gibbonstime}) fails to admit any 
``near-horizon limit,'' leading to the singular spacetime.   

The optimal way to capture the conformal structure of the spacetime is 
the introduction of the  conformal time 
\begin{align}
\label{}
\eta  \equiv -\frac{(n-3)}{(n-2)h } e^{-(n-2)h t/(n-3)}\,, \qquad 
-\infty <\eta <0 \,. 
\end{align}
In terms of $\eta$, 
the solution (\ref{Gibbonstime}) can be written in the form conformal
to the Gibbons solution as 
\begin{align}
\label{}
\D s^2 =&a^2 (\eta ) \left( - e^{-H_{\rm G}}\D \eta^2+
e^{H_{\rm G}/(n-3)}h_{IJ}\D y^I \D y^J\right)\,,\notag \\
\phi=& \sqrt{\frac{-\epsilon(n-2)}{4(n-3)\kappa_n}}\Bigl(H_{\rm G}+2(n-3)\ln a(\eta) \Bigr)\,,
\end{align}
where 
\begin{align}
\label{}
a(\eta ) \equiv \left[\frac{n-3}{-(n-2)h\eta}\right]^{1/(n-2)}\,. 
\end{align}
Let us consider for simplicity the single point source $H_{\rm G}(r)=M/r^{n-3}$ on the flat space 
$h_{IJ}\D y^I \D y^J=\D r^2+r^2 \D \Omega_{n-2}^2$. 
The radial null geodesic equations are integrated once to give 
\begin{align}
\label{}
k^\mu =\frac{e^{H_{\rm G}(r)}}{a^2(\eta)} \left(\frac{\partial}{\partial \eta}\right)^\mu \pm
 \frac{e^{(n-4)H_{\rm G}(r)/[2(n-3)]}}{a^2(\eta)}\left(\frac{\partial}{\partial r}\right)^\mu \,.
\end{align}
Along the geodesics, we have $\eta(r) =\pm \int^r e^{(n-2)H_{\rm G}(r)/[2(n-3)]}\D r$, thereby 
\begin{align}
\label{}
\lambda =\pm \int ^r a^2(\eta(r))e^{-(n-4)H_{\rm G}(r)/[2(n-3)]}\D r \,.
\end{align}

The Ricci scalar reads
\begin{align}
\label{}
R=&\frac{1}{4(n-2)a(\eta)^2}\left(\frac{12(n-1)}{\eta^2}e^{M/r^{n-3}}
-\frac{(n-2)^2(n-3)M^2}{r^{2(n-2)}} e^{-M/[(n-3)r^{n-3}]}
\right)\,.
\end{align}
The $k^\mu$ component of the Ricci tensor is 
\begin{align}
\label{}
R_{\mu\nu}k^\mu k^\nu=-\frac{n-3}{4(n-2)a(\eta)^4} \left(\frac{2}{\eta} e^{M/r^{n-3}}
+\frac{(n-2)M}{r^{n-2}} e^{(n-4)M/[2(n-3)r^{n-3}]}
\right)^2 \,.
\end{align}
The solution is therefore singular both at $\eta = -\infty$ where $a(\eta)\to 0$ and  
at $\eta = 0$ where $a(\eta)\to \infty$, as well as at $r=0$.

\end{document}